%% file: main.tex
\documentclass[11pt]{article}
\usepackage{jheppub}

\usepackage{amsmath}
\usepackage{amssymb}
\usepackage{amsthm}
\usepackage{graphicx}
\usepackage{subcaption}

\newtheorem{lemma}{Lemma}

\DeclareMathOperator{\Li}{Li}

\newcommand{\jint}[1]{I_{#1}}

\title{Residues on Permanent Pinches: Finite Integrals and Leading Divergences}
\author{Dimitri Corradini${}^1$,}%
\author{Cristian~Vergu${}^{1}$,}%
\author{Shun-Qing Zhang${}^1$}%
\emailAdd{dcorradi@mpp.mpg.de}
\emailAdd{c.vergu@gmail.com}
\emailAdd{sqzhang@mpp.mpg.de}
\affiliation{\small $^{1}$ Max-Planck-Institut f\"ur Physik, \\ Werner-Heisenberg-Institut,\\ Boltzmannstr.~8, 85748 Garching, Germany}

\abstract{We study infrared (IR) divergences from the point of view of permanent pinches.  We show how canceling permanent pinches allows us to form finite linear combinations of integrals.  We further show how to compute scheme-independent leading IR divergences without introducing a regulator, by integrating certain residue forms along desingularizations of permanent pinch varieties.  Our methods work for massless and mixed massive-massless integrals, for non-planar integrals with arbitrary numerators as well as for integrals with higher (integer) powers of propagators. We use this analysis to determine leading singular terms in some examples of box integrals. We further study examples of finite integrals at higher loops, and in particular provide a basis for finite non-evanescent integrals in a non-planar two-loop five-point topology.}

\preprint{MPP-2026-131}

\begin{document}
\maketitle
\newpage
\input{introduction}
\input{simple-examples}
\input{residue}

\input{double-box}
\input{thanks}

\appendix
\input{higher-residues}

\input{spherical-integral}
\input{nonplpent}

\bibliographystyle{jhep}
\bibliography{locally-finite}

\end{document}

%% file: introduction.tex
\section{Introduction}
\label{sec:intro}

Singularities of Feynman integrals are described in general by a set of solutions of the Landau equations (see ref.~\cite{Landau:1959fi}).  A usual Landau singularity occurs for codimension one (or higher) sub-varieties of the external kinematic space.  Said differently, one needs to adjust the external kinematics for a singularity to occur.

In contrast, infrared (IR) and ultraviolet (UV) divergences can be thought of as singularities of Feynman integrals which occur for \emph{all} values of external kinematics.  This perspective has been presented in an old paper~\cite{Kinoshita:1962ur} by Kinoshita, where the IR divergences are called ``mass singularities'' since they arise for generic external kinematics when particles become massless.

To indicate all such divergent regions, we will instead borrow the terminology \emph{permanent pinch} for such singularities from a paper~\cite{Boyling1968} by Boyling. This  illustrates the fact that for IR (or UV) singularities to occur, the contour of integration has to always be pinched, i.e.\ approached by poles on two or more converging sides, or at a boundary (or corner) of integration.  As described in textbook approaches to Landau singularities (see ref.~\cite{Eden1966}), if the contour is not pinched it can be moved to avoid the singularities and the integral is non-singular there.  For UV singularities, the pinching of the contour happens at infinity.  For IR singularities, permanent pinch Landau diagrams are very restricted topologically.  Indeed, one can see on explicit examples that they have a jet plus soft topology as is known from IR factorization properties (see refs.~\cite{PhysRevD.19.1250, PhysRevD.28.860}).

Solutions of permanent pinch equations define sub-varieties in the space of loop integration variables where divergences can arise. It is owing to these divergent regions that integrations require regularization, and thus develop poles of various orders in the parameter $\epsilon$ when performed in $D=d-2\epsilon$ dimensions, with $d$ integer (or logarithms of the masses in mass regularization~\cite{Kinoshita:1962ur}). In order to characterize, study and possibly subtract this divergent behavior, many different methods have been proposed and applied.

In momentum space, dimension shifts and
integration-by-parts identities allow
one to trade divergent integrals for quasi-finite ones, whose divergences
are isolated in explicit prefactors: this is exploited in
\cite{vonManteuffel:2014qoa} to construct quasi-finite bases of master
integrals, applied to massless form factors in
\cite{vonManteuffel:2015gxa}.  In this approach, finiteness is
achieved by relating different integrands, rather than through a
cancellation that is local on the integration contour.

The subtraction of divergences at the level of the integrand has its
origin in ultraviolet renormalization, where the BPHZ forest formula
\cite{Bogoliubov:1957gp, Hepp:1966eg, Zimmermann:1969jj} provides local
counterterms for each divergent subgraph; its extension to the infrared
divergences of Euclidean integrals is given by the $R^*$ operation
\cite{Chetyrkin:1982nn, Chetyrkin:1984xa, Smirnov:1985yck, Herzog:2017bjx}. In our language, these arise from purely soft permanent pinches; the collinear permanent pinches do not occur in Euclidean signature.  In ref.~\cite{brown2017feynmanamplitudescosmicgalois} a Hopf algebra of ``motic graphs'' has been described which, through the $R^*$ operation, can be shown (see ref.~\cite{Beekveldt:2020kzk}) to be relevant to Euclidean (non-physical) IR singularities. For physical cross sections, where infrared singularities cancel between
real and virtual corrections \cite{Kinoshita:1962ur, Lee:1964is},
subtraction schemes implement this cancellation through local counterterms
in phase space, beginning at NLO with \cite{Ellis:1980wv, Frixione:1995ms,
Catani:1996vz} and extended to NNLO in a variety of schemes (see
e.g.~\cite{Gehrmann-DeRidder:2005btv, Catani:2007vq,
Magnea:2018hab}). More recently, \cite{Capatti:2020xjc} introduces an approach that makes the cancellation of real and virtual contributions manifest locally at the level of cross-sections by defining them in terms of cuts of a given set of diagrams. Examples of counterterms introduced directly at the level of the loop integrals include \cite{Nagy:2003qn, Becker:2010ng}, which introduce subtraction methods suitable for numerical evaluation of one-loop QCD amplitudes, as well as the construction of locally finite
two-loop amplitudes from the universality of their infrared structure in
\cite{Anastasiou:2018rib,
Anastasiou:2020sdt,
Anastasiou:2024xvk, Anastasiou:2025cvy, Anastasiou:2026kpm}. In \cite{Georgoudis:2026han}, similar local counterterms are applied to the computation of pseudo-evanescent integrals: finite or divergent integrals whose integrand vanishes
identically in strictly four dimensions.

Another perspective is that of tropical geometry, based on the observation that the structure of many of the divergences that appear in integrations can be described in parameter space by the stratification of facets of the Newton polytope associated with a given integrand. This approach was first introduced in \cite{Pak:2010pt}, in the context of the determination of asymptotic regions for the expansion of integrals, necessary for their evaluation via the Method of Regions. Further work has focused on cases whose divergence is not directly encoded by tropicalization \cite{Gardi:2024axt, Jones:2024mfg}, as well as on the study of general graph-theoretical prescriptions for infrared regions \cite{Gardi:2022khw, Ma:2025emu, Ma:2026pjx}. Another approach to the characterization of divergent integrals in tropical geometry is described in \cite{Arkani-Hamed:2022cqe}. In \cite{Salvatori:2024nva}, the author introduces a subtraction procedure for integrals defined through asymptotic expansions in the regions defined by the tropical polytope. In a recent follow-up paper \cite{Giroux:2026tgd}, this method is implemented in a computer program in order to perform efficient symbolic integration. 

A different line of research involves the determination of a set of integrals with improved infrared properties for a given topology, with direct application, e.g., to the selection of bases of master integrals that may simplify integrations performed via the method of differential equations \cite{DeAngelis:2025agn}, as well as to fast numerical evaluation through specialized software \cite{Borinsky:2020rqs, Borinsky:2023nhb, trillo}. A notable predecessor to this line can be found in the discussion of finite dual-conformal invariant integrals contained in \cite{Arkani-Hamed:2010pyv}. In \cite{Wasser:2018qvj, Henn:2020lye} a method is introduced to classify a suitable set of integrals according to their infrared properties by expanding them in all soft and collinear regions and identifying combinations for which the corresponding divergences cancel. This procedure is applied in \cite{Henn:2019rmi} in order to select a particular basis, with a maximum order of divergence of $\epsilon^{-2}$, for master integrals in the computation of a contribution to the four-loop light-like cusp anomalous dimension in QCD. In \cite{Gambuti:2023eqh}, the authors describe an algorithm for determining the most general set of numerators that render integration convergent for a given topology, based on expansions in a parameterization around surfaces defined by the solutions of the Landau equations. This method is used in \cite{Figueiredo:2026ksz} to determine the set of such numerators in the case of massless planar \textit{pentabox} integrals.
The construction of identifying finite numerators has also been studied in Lee-Pomeransky representation, through a description based on tropical geometry, \cite{delaCruz:2024xsm} and
in the context of Loop-Tree Duality \cite{Dhani:2026cxx}.  Classification according to finiteness has been discussed in ref.~\cite{Bourjaily:2017wjl}, but there it was noted that separating finite and divergent integrals would complicate the form of some of their results.

In this paper, we will focus on the IR divergences in momentum space representation and show how a localization of integrands via residues on the permanent pinch varieties can elucidate some points regarding the structure of poles in the dimensional regulator and the corresponding subtraction schemes. In particular, we formulate the following conjecture: it is possible to extract the leading divergence of a Feynman integral from the ``pinch kinematics'' by taking a certain kind of (Leray) residues to obtain a differential form and by parameterizing the (real points) of the pinch kinematics to obtain an integration cycle.

More precisely, there is an operation which takes a pair $[\omega], [\gamma]$ of cohomology and homology class to a pair $[\omega_p], [\gamma_p]$ for any permanent pinch $p$.  Then, if $p$ is maximal (its corresponding Landau graph is not contained in another Landau graph of a permanent pinch), there is a contribution to the divergence in $\epsilon$ given by $\int_{\gamma_p} \omega_p$ (which is finite for $p$ maximal).  To obtain the leading divergence one should sum over maximal permanent pinches. We present below a few examples where this conjecture is satisfied. Note that the leading divergence, as defined above, will in general also contain subleading poles of different orders in $\epsilon$. More strongly, knowledge of the structure of inclusion between pinches, analogous to the lattice structure of the tropical polytope, should allow one to compute all contributions to a given integral, including all subleading poles in $\epsilon$, purely in terms of combinations of convergent integrals defined in integer dimensions on the pinch loci\footnote{Note that performing this procedure in practice requires additional prescriptions for the computation of consecutive residues, corresponding to the choice of a subtraction scheme}.

If this conjecture is true, it becomes possible to manipulate divergent integrals consistently (without using dimensional regularization) as long as the final result is finite\footnote{In fact, even the leading UV (as well as IR) divergences are well defined, which would allow us to compute anomalous dimensions, for example.}. In principle, this method allows us to distinguish between an integral which is \emph{convergent} and one for which the singularities cancel between different regions. Concretely, given a specific topology, we are able to identify numerators that correspond to locally finite integrals, i.e.\ integrals that are manifestly convergent, by solving linear systems in terms of residues on all pinch loci. As any such numerator is taken with respect to the same labeling, we emphasize that no momentum routing ambiguities occur (even for non-planar integrals).

\begin{figure}[hbtp]
   \centering
   \includegraphics[width=0.3\textwidth]{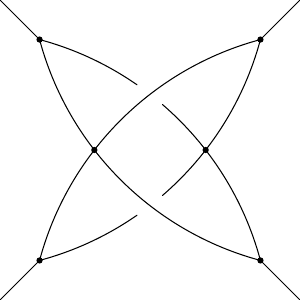}
   \caption{Three-loop example of a crown-type pinch}
   \label{fig:landshoff}
\end{figure}

Owing to known properties of the structure of permanent pinches, we will, perhaps surprisingly, never quite really need to solve Landau equations in practice. In fact, as we check explicitly in multiple instances, it appears to be sufficient in general to subtract a minimal subset of all divergences, which in a sense encode the whole structure of the singularities of a diagram, in order to achieve a complete subtraction of a given integrand to a corresponding locally finite one. These results, while presented here as heuristic, are in line with the description in~\cite{Arkani-Hamed:2022cqe}. In particular, when restricting to massless propagators, this minimal subset has a simple presentation in terms of specific contracted graph topologies: these are the so-called \textit{tadpoles} and scaleless \textit{bubbles}, as well as more exotic ones that first arise at higher loops and in some cases do not correspond to facets of the tropical polytope~\cite{Jones:2024mfg}, which we refer to as \textit{crown} or Landshoff type~\cite{PhysRevD.10.1024} (see fig.~\ref{fig:landshoff}). In any case, these latter will not appear in the simple topologies described in the present work.

\begin{figure}[hbtp]
     \centering
\includegraphics[width=0.2\textwidth]{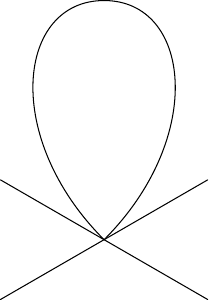}
\caption{One-loop example of a tadpole pinch}
\end{figure}

\begin{figure}[hbtp]
     \centering
\includegraphics[width=0.35\textwidth]{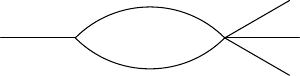}
\caption{One-loop example of a (scaleless) bubble pinch}
\label{fig:bubble}
\end{figure}

Once these regions are known for a given set of diagrams, we can extract the residue of integrand forms on them and use it to determine cancellations and divergent terms. As a first step in the more general program regarding the application of this procedure, we will mostly make use of the prototype example of one-loop box integrals, along with their subgraphs, in various massive and massless combinations, with unit or higher denominator powers and in different numbers of dimensions, to show certain central aspects of such calculations that already appear in these simple cases.  In sec.~\ref{sec:simple-examples}, we study some simple cases of box integrals with or without masses, and show how one can obtain expressions for their singularities from an analysis of their permanent pinches. In sec.~\ref{sec:residues} we aim to show explicitly the importance of the residue approach, as compared to the more common power-counting in terms of specific parameterizations. We describe explicitly some four- and five-dimensional cases where power divergences appear because of the presence of higher-order denominators, which possess some IR properties that would perhaps seem counterintuitive from a transverse-coordinate perspective. In sec.~\ref{sec:squared-tadpoles} we focus in particular on the computation of residues on configurations corresponding to tadpole pinches. In sec.~\ref{sec:two-loop-example}, we provide some examples of the application of this method at two loops and higher. In appendix~\ref{sec:higher-residues} we present some results from the theory of residues.

%% file: simple-examples.tex
\section{A few simple examples}
In this section, we introduce some examples aimed at familiarizing the reader with the core aspects of the procedure involved in our analysis of singularities: identifying permanent pinches that appear in a given topology, studying the behavior of the integrand on the associated locus in loop momenta, and eliminating it by subtracting subsector integrals. To this end, we purposely limit ourselves to examples for which the application of the residue machinery introduced in later sections can be avoided. 
\label{sec:simple-examples}
\subsection{Three- and two-mass boxes}
\begin{figure}
  \centering
  \includegraphics{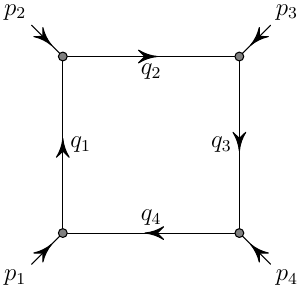}
  \caption{The kinematics of the box integral.}
  \label{fig:box}
\end{figure}

Consider the example of the three-mass box in four dimensions with external momenta $p_1, \dotsc, p_4$ and $p_1^2 = 0$, $p_i^2 = m_i^2$ for $i = 2, 3, 4$.  We label the internal momenta by $q_1, \dotsc, q_4$ with momentum conservation conditions $p_1 = q_1 - q_4$, $p_2 = q_2 - q_1$, $p_3 = q_3 - q_2$ and $p_4 = q_4 - q_3$, as detailed in fig.~\ref{fig:box}.  This labeling will be used throughout this section.

This integral is IR-divergent, which can be seen by noticing that it has a permanent pinch; the Landau equations are satisfied for $q_1 = z p_1$ and $q_4 = -(1 - z) p_1$, irrespective of the external kinematics.  Also note that the Landau equations do not have a unique solution, but a whole family of solutions parameterized by the momentum fraction $z$. This solution identifies a collinear configuration, and it can be represented through a contracted diagram with the topology of a scaleless bubble, as in fig.~\ref{fig:bubble}.

When the Landau equations are satisfied, we can solve for the remaining momenta in terms of the momentum fractions.  We have
\begin{gather}
  q_2 = p_2 + q_1, \qquad
  q_2^2 = (1 - z) m_2^2 + z s_{12}, \\
  q_3 = q_4 - p_4, \qquad
  q_3^2 = (1 - z) s_{23} + z m_4^2,
\end{gather}
where we have used the notation $s_{i j} = (p_i + p_j)^2$.

We can now show by a straightforward partial-fractioning that
\begin{equation}
  \label{eq:three-mass-identity}
  \frac {s_{12} s_{23} - m_2^2 m_4^2}{q_2^2 q_3^2} - \frac {s_{12} - m_2^2}{q_2^2} - \frac {s_{23} - m_4^2}{q_3^2} = 0.
\end{equation}
This suggests that this bubble permanent pinch in the three-mass box integral can be canceled by two triangle integrals with appropriate coefficients.  As we will discuss in sec.~\ref{sec:residues}, tadpoles do not contribute to IR divergences of this topology in four dimensions. As such, this is the only permanent pinch that we need to consider, which suggests that the linear combination that cancels it should be IR-finite.

This can be explicitly checked as follows.  We will use the explicit expressions from ref.~\cite[sec.~4.4]{Bern:1993kr} for the three-mass box with massless propagators around four dimensions at energy scale $\mu$:
\begin{multline}
  \label{eq:three-mass-box}
  I_4^D(0, m_2^2, m_3^2, m_4^2; s_{12}, s_{23}) =
  \frac {\mu^{2 \epsilon} i \pi^{\frac{D}{2}}}{s_{12} s_{23} - m_2^2 m_4^2} \Bigl\lbrace\\
  2 \epsilon^{-2} \Bigl[
      (-s_{12})^{-\epsilon} +
      (-s_{23})^{-\epsilon} -
      (-m_2^2)^{-\epsilon} -
      (-m_3^2)^{-\epsilon} -
      (-m_4^2)^{-\epsilon}\Bigr] + \\
      \epsilon^{-2} \Bigl[
      (-m_2^2)^{-\epsilon} (-m_3^2)^{-\epsilon} (-s_{23})^{\epsilon} +
      (-m_3^2)^{-\epsilon} (-m_4^2)^{-\epsilon} (-s_{12})^{\epsilon}\Bigr] + \\
  -2 \Li_2(1 - \frac {m_2^2}{s_{12}})
  -2 \Li_2(1 - \frac {m_4^2}{s_{23}})
  + 2 \Li_2(1 - \frac {m_2^2 m_4^2}{s_{12} s_{23}})
  - \log^2 \frac {s_{12}}{s_{23}} +
  \mathcal{O}(\epsilon) \Bigr\rbrace.
\end{multline}
We emphasize that even though it naively appears that the expansion in a Laurent series of $I_4^D$ of this result starts at $\epsilon^{-2}$, the corresponding term actually cancels, and the expansion starts at order $\epsilon^{-1}$. As we will see in examples below, this can be tied to the absence of triangle permanent pinches.  Moreover, the leading IR divergence reads
\begin{equation}
    \label{eq:three-mass-pole}
    \frac{i \pi^2}{\epsilon} \frac{1}{s_{12} s_{23} - m_2^2 m_4^2} \bigl(-\log (-s_{12}) - \log(-s_{23}) + \log(-m_2^2) + \log(-m_4^2)\bigr).
\end{equation}
This divergence can be predicted, up to the overall factor $-i \pi^2/\epsilon$,\footnote{As we will see in sec.~\ref{sec:residue-other-dims}, the complete normalization also involves a trivial integral over a sphere, providing additional numerical factors. In this initial section we only aim to show that this general procedure provides the correct kinematics dependence, and thus neglect these overall constants.} by integrating the integrand in the pinch kinematics over $z \in [0, 1]$, since
\begin{multline}
    \int_0^1 \frac{d z}{((1 - z) m_2^2 + z s_{12}) ((1 - z) s_{23} + z m_4^2)} =\\
    \frac{1}{s_{12} s_{23} - m_2^2 m_4^2} \bigl(\log (-s_{12}) + \log(-s_{23}) - \log(-m_2^2) - \log(-m_4^2)\bigr).
\end{multline}

We also need the two-mass triangle integrals with massless propagators around four dimensions (see ref.~\cite{Beenakker:2002nc}):
\begin{equation}
\begin{aligned}
  \label{eq:two-mass-triangle}
  I_3^D(0, m_2^2, m_3^2) &= \frac {\mu^{2 \epsilon} i \pi^{\frac{D}{2}}}{\epsilon^2} \frac 1 {m_2^2 - m_3^2} \Bigl((-m_2^2)^{-\epsilon} - (-m_3^2)^{-\epsilon}\Bigr)  \\&=
  \frac 1 {m_2^2 - m_3^2} \Bigl\lbrace
  \frac 1 \epsilon \log \frac {m_3^2}{m_2^2} +
  \frac 1 2 \Bigl[
  \log^2 \frac {-m_2^2}{\mu^2} -
  \log^2 \frac {-m_3^2}{\mu^2}
  \Bigr] +
  \mathcal{O}(\epsilon) \Bigr\rbrace.
\end{aligned}
\end{equation}
The same expressions can be extracted from ref.~\cite{Ellis:2007qk}.

Computing
\begin{equation}
\begin{aligned}
  &(s_{12} s_{23} - m_2^2 m_4^2) I_4^D(0, m_2^2, m_3^2, m_4^2; s_{12}, s_{23}) \\&-
  (s_{12} - m_2^2) I_3^D(0, m_2^2, s_{12}) -
  (s_{23} - m_4^2) I_3^D(0, m_4^2, s_{23})
\end{aligned}
\end{equation}
we find for the divergent part
\begin{multline}
  \mu^{2 \epsilon} \epsilon^{-2} i \pi^{\frac{D}{2}} \Bigl(
  2 (-s_{12})^{-\epsilon} +
  2 (-s_{23})^{-\epsilon} -
  2 (-m_2^2)^{-\epsilon} -
  2 (-m_3^2)^{-\epsilon} -
  2 (-m_4^2)^{-\epsilon} + \\
  (-m_2^2)^{-\epsilon} (-m_3^2)^{-\epsilon} (-s_{23})^{\epsilon} +
  (-m_3^2)^{-\epsilon} (-m_4^2)^{-\epsilon} (-s_{12})^{\epsilon} - \\
  ((-s_{12})^{-\epsilon} - (-m_2^2)^{-\epsilon}) -
  ((-s_{23})^{-\epsilon} - (-m_4^2)^{-\epsilon})\Bigr) = \\
  i \pi^2 \Bigl(\frac 1 2 \log^2 (-s_{12}) -
  \frac 1 2 \log^2 (-m_3^2) -
  \frac 1 2 \log^2 (-m_4^2) + \\
  \log (-m_3^2) \log (-m_4^2) -
  \log (-m_3^2) \log (-s_{12}) -
  \log (-m_4^2) \log (-s_{12}) + \\
  (s_{12} \to s_{23}, m_2 \to m_4)\Bigr) + \mathcal{O}(\epsilon).
\end{multline}

In the end, we have
\begin{multline}
  \lim_{D \to 4} \Bigl(
  (s_{12} s_{23} - m_2^2 m_4^2) I_4^D(0, m_2^2, m_3^2, m_4^2; s_{12}, s_{23}) - \\
  (s_{12} - m_2^2) I_3^D(0, m_2^2, s_{12}) -
  (s_{23} - m_4^2) I_3^D(0, m_4^2, s_{23})
  \Bigr) = \\ i \pi^2 \Bigl(
  -2 \Li_2(1 - \frac {m_2^2}{s_{12}})
  -2 \Li_2(1 - \frac {m_4^2}{s_{23}})
  + 2 \Li_2(1 - \frac {m_2^2 m_4^2}{s_{12} s_{23}})
  - \log^2 \frac {s_{12}}{s_{23}} + \\
  \frac 1 2 \log^2 (-s_{12}) -
  \frac 1 2 \log^2 (-m_3^2) -
  \frac 1 2 \log^2 (-m_4^2) + \\
  \log (-m_3^2) \log (-m_4^2) -
  \log (-m_3^2) \log (-s_{12}) -
  \log (-m_4^2) \log (-s_{12}) + \\
  \frac 1 2 \log^2 (-s_{23}) -
  \frac 1 2 \log^2 (-m_3^2) -
  \frac 1 2 \log^2 (-m_2^2) + \\
  \log (-m_3^2) \log (-m_2^2) -
  \log (-m_3^2) \log (-s_{23}) -
  \log (-m_2^2) \log (-s_{23})\Bigr).
\end{multline}

A variant of this box integral, where the propagators corresponding to momenta $q_2$ and $q_3$ have masses $\mu_2$ and $\mu_3$ respectively is just as easily analyzed.  Indeed, we find
\begin{equation}
  \label{eq:three-mass-bow-two-mass-propagators}
  \frac {s_{12} s_{23} - s_{12} \mu_3^2 - s_{23} \mu_2^2 + m_2^2 \mu_3^2 + m_4^2 \mu_2^2 - m_2^2 m_4^2}{(q_2^2 - \mu_2^2) (q_3^2 - \mu_3^2)} -
  \frac {s_{12} - m_2^2}{q_2^2 - \mu_2^2} -
  \frac {s_{23} - m_4^2}{q_3^2 - \mu_3^2} = 0
\end{equation}
for $q_1 = z p_1$ and $q_4 = -(1 - z) p_1$.  We note that the somewhat complicated integral prefactor $s_{12} s_{23} - s_{12} \mu_3^2 - s_{23} \mu_2^2 + m_2^2 \mu_3^2 + m_4^2 \mu_2^2 - m_2^2 m_4^2$ is obtained effortlessly (compare with ref.~\cite[eq.~4.38]{Ellis:2007qk}).  We also note that the leading singularity of the two-mass triangles with one massive propagator remains unchanged with respect to the two-mass triangle with all massless propagators (compare eqs.~4.6 and 4.8 in ref.~\cite{Ellis:2007qk}).

This cancels the bubble permanent pinch and the analogous linear combination of integrals is IR-finite.  In fact, this linear combination is even nicer than the one with massless $q_2$ and $q_3$ propagators, since in the massless case tadpole permanent pinches remain.

For example, the permanent pinch corresponding to the $q_2$ tadpole is given by a Landau equation $\alpha_2 q_2 = 0$ so we obtain $q_2 = 0$, which is a soft singularity.  This singularity is integrable above two dimensions but in two dimensions it leads to IR singularities.\footnote{One is used to setting the tadpole integral to zero in dimensional regularization.  The argument for this goes as follows (see ref.~\cite{Collins_1984}).  We want the dimensionally-regularized integral to have the usual properties of linearity and change of coordinates.  Take:
\[
  I^{(0)} = \int \frac {d^D p}{p^2}
\]
for a massless tadpole.  Make a change of coordinates $p \to s p$, where $s$ is a scalar we have $d^D s p = s^D d^D p$ and $p^2 \to s^2 p^2$ so the integral becomes
\[
\int \frac {d^D p}{p^2} = \int s^{D - 2} \frac {d^D p}{p^2} = s^{D - 2} \int \frac {d^D p}{p^2}.
\]
Now, \emph{if} $D \neq 2$ we can conclude that the tadpole vanishes.  Exactly when $D = 2$ this argument fails to show that the integral vanishes.}  The same happens for squared propagators in four dimensions.

\begin{figure}
     \centering
     \begin{subfigure}[m]{0.3\textwidth}
         \centering
         \includegraphics[width=\textwidth]{figures/bubble}
       \end{subfigure}
     \hfil
     \begin{subfigure}[m]{0.3\textwidth}
         \centering
         \includegraphics[width=\textwidth]{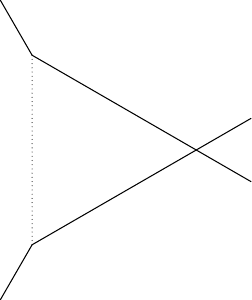}
     \end{subfigure}
     \caption{The permanent pinches of the massless box integral (up to symmetry).  The bubble permanent pinch has pinch kinematics where the momenta in the bubble are collinear to the external momentum.  The triangle permanent pinch has the dotted momentum equal to zero (soft momentum).  In turn, this forces the adjacent loop momenta to be equal to the external momenta and therefore to be on-shell.}
     \label{fig:massless-box-pinches}
\end{figure}
By the same token, it is possible to determine divergent terms in the so-called \textit{two-mass easy} configuration, corresponding to $p_1^2 = p_3^2 = 0$, $p_2^2 = m_2^2$, $p_4^2 = m_4^2$ in the labeling of fig.~\ref{fig:box}. There are now two bubble permanent pinches, one for each massless external leg. On the collinear configuration $q_1 = z p_1$, $q_4 = -(1 - z) p_1$ we find
\begin{gather}
  q_2^2 = (1 - z) m_2^2 + z s_{12}, \qquad
  q_3^2 = (1 - z) s_{23} + z m_4^2,
\end{gather}
exactly as for the three-mass box: since the identity in eq.~\eqref{eq:three-mass-identity} never involved $m_3^2$, it applies verbatim and fixes the coefficients of the triangles $I_3^D(0, m_2^2, s_{12})$ and $I_3^D(0, m_4^2, s_{23})$.  On the second collinear configuration, $q_2 = -\bar z p_3$ and $q_3 = (1 - \bar z) p_3$, the surviving propagators are
\begin{gather}
  q_1^2 = (1 - \bar z) m_2^2 + \bar z s_{23}, \qquad
  q_4^2 = (1 - \bar z) s_{12} + \bar z m_4^2,
\end{gather}
and the same identity with $s_{12} \leftrightarrow s_{23}$ fixes the coefficients of $I_3^D(0, m_2^2, s_{23})$ and $I_3^D(0, m_4^2, s_{12})$.  The combination
\begin{multline}
  \label{eq:two-mass-easy-finite}
  (s_{12} s_{23} - m_2^2 m_4^2) I_4^D(0, m_2^2, 0, m_4^2; s_{12}, s_{23}) -
  (s_{12} - m_2^2) I_3^D(0, m_2^2, s_{12}) - \\
  (s_{23} - m_4^2) I_3^D(0, m_4^2, s_{23}) -
  (s_{23} - m_2^2) I_3^D(0, m_2^2, s_{23}) -
  (s_{12} - m_4^2) I_3^D(0, m_4^2, s_{12})
\end{multline}
is therefore free of collinear permanent pinches and IR-finite.

Each collinear pinch contributes its own $z$-integral to the leading divergence, and the two integrals coincide, so the leading IR divergence is now \emph{twice} the single-pinch result:
\begin{multline}
    I_4^D(0, m_2^2, 0, m_4^2; s_{12}, s_{23}) = \frac{i \pi^2}{\epsilon} \, \frac{2}{s_{12} s_{23} - m_2^2 m_4^2} \\ \times \bigl(-\log (-s_{12}) - \log(-s_{23}) + \log(-m_2^2) + \log(-m_4^2)\bigr) + \mathcal{O}(\epsilon^0),
\end{multline}
which matches the known result for the $\frac 1 \epsilon$ pole of the two-mass easy box (see e.g.\ \cite[eq.~4.20]{Ellis:2007qk}).
\subsection{Massless box}
Let us now consider the massless box integral, with the same labeling of internal momenta as in fig.~\ref{fig:box}.  This integral has many more permanent pinches and therefore has a more involved IR structure compared to examples with masses.  Neglecting tadpoles, it has the two types of permanent pinches given in fig.~\ref{fig:massless-box-pinches}, up to symmetries.  It is convenient to introduce the notation
\begin{equation}
  I(\nu_1, \nu_2, \nu_3, \nu_4) = \int \frac {d^D q_1}{(q_1^2)^{\nu_1} (q_2^2)^{\nu_2} (q_3^2)^{\nu_3} (q_4^2)^{\nu_4}},
\end{equation}
where the momenta $q_2$, $q_3$, $q_4$ are determined by $q_1$ and the external momenta through the momentum conservation conditions above.

Just as before, our aim is to introduce a combination of subsector integrals that causes the expression to vanish on the kinematics defined by such pinches.  Consider first the bubble permanent pinch associated with $p_1$ (left in fig.~\ref{fig:massless-box-pinches}), $q_1 = z p_1$ and $q_4 = -(1 - z) p_1$.  The two remaining propagators evaluate to $q_2^2 = z s_{12}$ and $q_3^2 = (1 - z) s_{23}$, and we have
\begin{equation}
  \frac 1 {q_2^2 q_3^2} =
  \frac 1 {z (1 - z) s_{1 2} s_{2 3}} =
  \frac 1 {s_{2 3}} \frac 1 {z s_{1 2}} + \frac 1 {s_{1 2}} \frac 1 {(1 - z) s_{2 3}} =
  \frac 1 {s_{2 3}} \frac 1 {q_2^2} + \frac 1 {s_{1 2}} \frac 1 {q_3^2}.
\end{equation}
The two triangles that share this pinch are those obtained by dropping one of the two off-shell propagators, $I(1, 1, 0, 1)$ and $I(1, 0, 1, 1)$, and the identity fixes their coefficients to $-\frac 1 {s_{23}}$ and $-\frac 1 {s_{12}}$ respectively.  Each of the four bubble permanent pinches yields an identity of this form, and each triangle shares two of them.  The resulting conditions are consistent: every triangle appears with the same coefficient in both identities in which it participates, and the result is
\begin{equation}
    I(1,1,1,1)-\frac{1}{s_{23}}\left(I(0,1,1,1)+I(1,1,0,1)\right)-\frac{1}{s_{12}}\left(I(1,0,1,1)+I(1,1,1,0)\right).
\label{eq:box-minus-triangles}
\end{equation}
We next consider the triangle permanent pinches.  For a soft $q_1$ the adjacent momenta are forced on-shell, $q_2 = p_2$ and $q_4 = -p_1$, and the single remaining off-shell propagator of the box evaluates to $q_3^2 = (p_1 + p_4)^2 = s_{23}$.  Among the triangles, precisely one shares this pinch, namely the one obtained by dropping that off-shell propagator, $I(1, 1, 0, 1)$, with residue $1$.  Since the coefficient of $I(1, 1, 0, 1)$ in eq.~\eqref{eq:box-minus-triangles} is $-\frac 1 {s_{23}}$, this pinch cancels as well, and the same happens for the remaining three pinches of this type.  The conditions imposed by the cancellation of triangle permanent pinches thus yield the \emph{same} solutions for the triangle coefficients as the cancellation of bubble permanent pinches. This can be seen as a manifestation of the claim made in sec.~\ref{sec:intro} regarding the minimal set of pinch regions relevant for subtractions.

Consider now the box integral with one internal mass:
\begin{equation}
    I_m(\nu_1, \nu_2, \nu_3, \nu_4) = \int \frac {d^D q_1}{(q_{1}^2)^{\nu_1} (q_{2}^2)^{\nu_2} (q_{3}^2 - m^2)^{\nu_3} (q_{4}^2)^{\nu_4}}.
\end{equation}
We first study the bubble permanent pinch where $q_1 = z p_1$ and $q_4 = -(1 - z) p_1$.  In this kinematics we have $q_2^2 = z s_{12}$ and $q_3^2 - m^2 = (1 - z) s_{23} - m^2$.  Then,
\begin{equation}
    \frac 1 {z s_{12} ((1 - z) s_{23} - m^2)} =
    \frac 1 {s_{12} (s_{23} - m^2)}\, \frac 1 z +
    \frac {s_{23}}{s_{12} (s_{23} - m^2)}\, \frac 1 {(1 - z) s_{23} - m^2}.
\end{equation}
Since the massive propagator cannot participate in a collinear pinch, the only other bubble permanent pinch is the one associated with $p_2$, where $q_1 = -z p_2$ and $q_2 = (1 - z) p_2$; there the surviving propagators are $q_4^2 = z s_{12}$ and $q_3^2 - m^2 = (1 - z) s_{23} - m^2$, and the identity above is reproduced with the roles of $q_2^2$ and $q_4^2$ exchanged.  In particular, the triangle $I_m(1, 1, 0, 1)$ is shared by both pinches and appears with the same coefficient in both identities.  This suggests that the linear combination
\begin{equation}
    \label{eq:box-1mass-finite}
    I_m(1, 1, 1, 1) -
    \frac{1}{s_{23} - m^2} I_m(1,1,0,1) -
    \frac{s_{23}}{s_{12} (s_{23} - m^2)} (I_m(1, 0, 1, 1) + I_m(1, 1, 1, 0))
\end{equation}
is finite. We have verified the finiteness with numerical integration software such as {\tt AMFlow}~\cite{Liu:2017jxz,Liu:2022chg,Huang:2026rjb}, \texttt{trillo}~\cite{trillo} and \texttt{pySecDec}~\cite{Borowka:2017idc,Heinrich:2023til}. Note that this combination and all the previous ones do not represent the unique solution for finite integrals in this topology. In particular, in following~\cite{Gambuti:2023eqh}, the most general ansatz for a finite integral in this topology should independently include the six Lorentz-invariant independent quantities available in this setting, as well as their products up to degree three in the loop momentum, with the latter bound enforcing UV-finiteness. We will demonstrate an example of a more general calculation in the case of a two-loop topology in sec.~\ref{sec:two-loop-example}.

So far, we have carried out computations in a somewhat naive way, simply picking some permanent pinch kinematics and evaluating the integrand (except for the on-shell propagators) at that point, imposing that it vanishes.  In the following, we will argue that the mathematically proper way to formulate this recipe is in terms of residues that localize the integral to the corresponding divergent locus (after performing the necessary blow-ups, i.e.\ changes of variables required to resolve potential singular configurations).

As a preview of the formalism developed in the following, we end this examples section with the box integral $I_m(3, 1, 1, 1)$, with one cubed propagator.  In an expansion around four dimensions it behaves as
\begin{equation}
    \begin{aligned}
&I_m(3, 1, 1, 1) = \epsilon^{-2} \frac {i \pi^2}{(s_{12})^3 (m^2 - s_{23})^5} \\&\times\Bigl(m^4 s_{12}^2 +
    6 m^4 s_{12} s_{23} +
    4 m^2 s_{12}^2 s_{23} +
    6 m^4 s_{23}^2 +
    6 m^2 s_{12} s_{23}^2 +
    s_{12}^2 s_{23}^2\Bigr) + \mathcal{O}(\epsilon^{-1}).
        \end{aligned}
\end{equation}
This highly nontrivial coefficient can be computed by taking the triangle residue, which corresponds to the only maximal permanent pinch in this topology. There are lower divergences, arising at $\mathcal{O}(\epsilon^{-1})$, which correspond to bubble or higher power tadpole permanent pinches.  The purely algebraic method for computing this triangle residue is described in the next section.

%% file: residue.tex
\section{Computing residues on permanent pinches}
\label{sec:residues}
As we have shown in examples in the previous section, permanent pinches define loci in the space of loop momenta which are associated to the emergence of singularities in the integration. To obtain the value of the corresponding pole in $\epsilon$, we need to perform an integration of some quantity describing the localization of the integrand form to this pinch locus. In particular, in order for the procedure to be well-defined, this should be independent of the parameterization of the pinch locus, as well as, more in general, holomorphic changes of variables in the integrand form that are locally invertible around the pinch surface. The natural candidate for such an object is the residue form of the integrand on the pinch surface\footnote{More precisely, the residue form is intrinsically defined only up to total derivatives. We will make use of this distinction in the following sec.~\ref{sec:cohomology}}. A summary of useful aspects of Leray's theory of residues is presented in appendix~\ref{sec:higher-residues}. As a short justification for this procedure, we note here that poles which are not simple, computed in terms of a Laurent expansion in variables parameterizing the directions transverse to a given pinch surface in the space of loop momenta, can always be rewritten as total derivatives, which evaluate to zero both in integer dimensions and in dimensional regularization. Note that even in this picture, the vanishing of higher-pole coefficients in power counting can still determine, in general, a good, coordinate-independent sufficient condition for the absence of divergences.

In this section, we provide examples on how to use residue computations to derive divergences and  finite integrals in circumstances in which power divergences appear in infrared regions, in particular, through propagators of higher power (which we will refer to as \textit{dotted} propagators) in one-loop topologies. Note that in general, at two or more loops these also appear in topologies with no dotted propagator.

\begin{figure}
  \centering
  \includegraphics{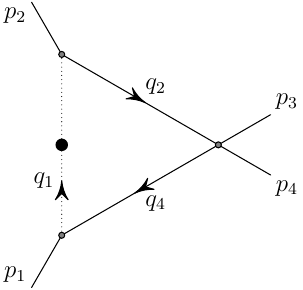}
  \caption{A triangle permanent pinch with a soft propagator (marked by dotted line).  We have also marked a squared propagator by a dot, since we will be concerned with taking a residue of a differential form with a pole of order two where the dotted propagator is singular.}
  \label{fig:triangle-pinch1}
\end{figure}
\subsection{Triangle integrals}
To start, we will describe how to compute a residue in a triangle integral with massless internal propagators, two massless external legs, for which we write $p_1 = \lambda_1 \tilde{\lambda}_1$, $p_2 = \lambda_2 \tilde{\lambda}_2$ in terms of spinor helicity variables, and one massive leg (which for us will consist of two massless propagators $p_3 + p_4$, see fig.~\ref{fig:triangle-pinch1}).  This will be even more interesting for the case when one of the internal propagators in the triangle is a squared propagator.
We find it easiest to take the residue in two steps; first we will take the residue in two of the propagators which form a bubble graph under contraction.\footnote{A similar calculation for the triangle, but in a different language, was done in ref.~\cite[sec.~3.5]{Bourjaily:2013mma}.}  For this purpose we will parameterize the kinematics in this bubble, whose momenta we denote by $q_2$ and $q_4$, oriented so that $p_1 + p_2 - q_2 + q_4 = 0$.  One can show that generically this on-shell kinematics is parameterized by
\begin{align}
  q_2 &= \frac 1 {1 - z w} (w \lambda_1 + \lambda_2) (-z \tilde{\lambda}_1 + \tilde{\lambda}_2), \\ 
  q_4 &= \frac 1 {1 - z w} (\lambda_1 + z \lambda_2) (-\tilde{\lambda}_1 + w \tilde{\lambda}_2),
\end{align}
where $z, w \in \mathbb{C}$.  More precisely, one could say that the (complexified) kinematics is parameterized by $\mathbb{P}^1 \times \mathbb{P}^1$ and $z, w$ are coordinates on affine patches on these $\mathbb{P}^1$.

We have\footnote{Given that $p^{1 \dot{1}} = p^0 + p^3$, $p^{1 \dot{2}} = p^1 - i p^2$, $p^{2 \dot{1}} = p^1 + i p^2$ and $p^{2 \dot{2}} = p^0 - p^3$, we obtain that $p^0 \wedge p^1 \wedge p^2 \wedge p^3 = \frac i 4 p^{1 \dot{1}} \wedge p^{1 \dot{2}} \wedge p^{2 \dot{1}} \wedge p^{2 \dot{2}}$.  Below we use the notation $d^4 p = d p^{1 \dot{1}} \wedge d p^{1 \dot{2}} \wedge d p^{2 \dot{1}} \wedge d p^{2 \dot{2}}$ but, as long as we only compare residues of different integrals, this normalization will not influence the results.  These conventions also imply that $p^2 = \det p^{\alpha \dot{\alpha}}$ and $(p + q)^2 = \langle \lambda_p \lambda_q\rangle [\tilde{\lambda}_p \tilde{\lambda}_q]$.}
\begin{equation}
  \operatorname{res}_{q_2^2 = q_4^2 = 0} \frac {d^4 q_2}{q_2^2 q_4^2} =
  \frac {d q_2^{2 \dot{1}} d q_2^{2 \dot{2}}}{\det \left(\begin{smallmatrix} q_2^{2 \dot{2}} & q_4^{2 \dot{2}} \\ -q_2^{2 \dot{1}} & -q_4^{2 \dot{1}}\end{smallmatrix}\right)} =
  -\frac {d z d w}{(1 - z w)^2},
\end{equation}
where we have used $p^2 = p^{1 \dot{1}} p^{2 \dot{2}} - p^{2 \dot{1}} p^{1 \dot{2}}$.  This measure lifts to a globally defined meromorphic two-form on $\mathbb{P}^1 \times \mathbb{P}^1$.

If we add an extra propagator $q_1^2 = (p_1 + q_4)^2 = s_{1 2} \frac {z w}{1 - z w}$, we obtain\footnote{One may wonder what is the interpretation of the pole at $z w = 1$.  By plugging back into the expressions for momenta $q_1, q_2$ in terms of $z, w$, we can see that the momentum components become infinite, so we are dealing with a UV kinematics. At generic kinematics, this pole at infinity, whose locus doesn't intersect the real slice of the integration space, doesn't correspond to a pinch to the integration contour. Thus the integral remains UV-finite, as power counting confirms.}
\begin{gather}
  \operatorname{res}_{q_2^2 = q_4^2 = 0} \frac {d^4 q_2}{q_1^2 q_2^2 q_4^2} = -\frac 1 {s_{1 2}} \frac {d z d w}{z w (1 - z w)}, \label{eq:triangle-trianglepinch} \\
  \operatorname{res}_{q_2^2 = q_4^2 = 0} \frac {d^4 q_2}{(q_1^2)^2 q_2^2 q_4^2} = -\frac 1 {s_{1 2}^2} \frac {d z d w}{z^2 w^2},
\end{gather}
where we have included the case of a squared propagator as well.

Taking the residues $z = 0$ and $w = 0$ localizes the kinematics to that of a triangle permanent pinch with $q_1 = 0$.  If the propagator isn't dotted, taking the residue amounts to replacing the momentum in the rest of the diagram by using momentum conservation, thus obtaining $(-s_{12})^{-1}$. However, in the case of a squared propagator we need to isolate the coefficient of $z w$ in a double expansion around $z = w = 0$.  For a triangle with a squared propagator we find zero.  If we include another propagator $q_3^2 = (p_4 - q_4)^2$, we find
\begin{equation}
  \operatorname{res}_{z = w = 0} \operatorname{res}_{q_2^2 = q_4^2 = 0} \frac {d^4 q_2}{(q_1^2)^2 q_2^2 q_4^2 q_3^2} = -\frac 1 {s_{1 2} s_{2 3}^2}.
\end{equation}

Since the residue for a triangle with squared propagator vanishes, the singularity of a box with squared propagator cannot be canceled by only adding a multiple of such a triangle. We can instead consider a triangle with numerator, such as
\begin{equation}
  \int \frac {d^4 q_2\; q_3^2}{(q_1^2)^2 q_2^2 q_4^2}.
  \label{eq:triangledoublenum}
\end{equation}
Taking a residue, we obtain
\begin{equation}
  \operatorname{res}_{z = w = 0} \operatorname{res}_{q_2^2 = q_4^2 = 0} \frac {d^4 q_2\; q_3^2}{(q_1^2)^2 q_2^2 q_4^2} = -\frac 1 {s_{1 2}^2} (s_{1 2} + 2 s_{2 3}).
\end{equation}

We find that the singularity of a box with squared propagator in the triangle pinch configuration we have been studying above can be canceled by the triangle~\eqref{eq:triangledoublenum} multiplied by a coefficient
\begin{equation}
  \frac {s_{12}}{s_{23}^2 (s_{12} + 2 s_{23})}.
\end{equation}

One may be worried about the potential singularity at $s_{12} + 2 s_{23} = 0$ which looks unphysical.  It turns out that the triangle integral~\eqref{eq:triangledoublenum} \emph{vanishes} when $s_{12} + 2 s_{23} = 0$ thus canceling this singularity.

\begin{figure}
  \centering
  \includegraphics{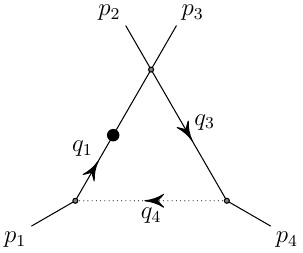}
  \caption{Another triangle permanent pinch with a soft propagator (marked by dotted line).  We have marked one of the propagators by a dot to indicate which propagator appears squared in the calculation.}
  \label{fig:triangle-pinch2}
\end{figure}

Let us now compute a triple residue in another triangle permanent pinch kinematics.  This time, the momentum going soft will not be squared as in the previous case.  The momenta in the triangle will be denoted by $q_1$, $q_3$, $q_4$ (the labeling is obtained by contracting the box kinematics) where $q_3 = q_1 - p_1 - p_4$ and $q_4 = q_1 - p_1$ (see fig~\ref{fig:triangle-pinch2}).  We want to compute
\begin{equation}
  \operatorname{res}_{q_1^2 = q_3^2 = q_4^2 = 0} \frac {d^4 q_1}{(q_1^2)^2 q_3^2 q_4^2}.
\end{equation}
We proceed as before, by computing the residue in $q_3^2$ and $q_4^2$ first. The calculation differs from the previous case, since now the kinematics is either
\begin{equation}
  q_3 = \lambda_4 (\tilde{\lambda} - \tilde{\lambda}_4), \qquad
  q_4 = \lambda_4 \tilde{\lambda},
\end{equation}
or
\begin{equation}
  q_3 = (\lambda - \lambda_4) \tilde{\lambda}_4, \qquad
  q_4 = \lambda \tilde{\lambda}_4.
\end{equation}
In this case, the on-shell space for $q_3^2 = q_4^2 = 0$ has two branches, the first parameterized by $\tilde{\lambda}$ and the second one by $\lambda$.  The soft region in the first branch is $\tilde{\lambda} = 0$ while it is $\lambda = 0$ for the second branch.  These can be seen as coordinates on $\mathbb{C}^2$, unlike $\mathbb{P}^1 \times \mathbb{P}^1$ in the previous case.

Taking the residues in the first branch, we find
\begin{equation}
  \operatorname{res}_{q_3^2 = q_4^2 = 0} \frac {d^4 q_1}{(q_1^2)^2 q_3^2 q_4^2} =
  \frac {d q_1^{2 \dot{1}} d q_1^{2 \dot{2}}}{(q_1^2)^2 \det \left(\begin{smallmatrix} q_3^{2 \dot{2}} & q_4^{2 \dot{2}} \\ -q_3^{2 \dot{1}} & -q_4^{2 \dot{1}}\end{smallmatrix}\right)} =
  \frac {d^2 \tilde{\lambda}}{\langle 1 4\rangle^2 [1 \tilde{\lambda}]^2 [\tilde{\lambda} 4]},
\end{equation}
where we have used $d q_3^{2 \dot{1}} d q_3^{2 \dot{2}} = (\lambda_4^2)^2 d^2 \tilde{\lambda}$, $q_1^2 = (q_4 + p_1)^2 = \langle 1 4\rangle [1 \tilde{\lambda}]$ and the determinant is $(\lambda_4^2)^2 [\tilde{\lambda} 4]$.  Using
\begin{equation}
  d^2 \tilde{\lambda} = \frac {d [\tilde{\lambda} a] d [\tilde{\lambda} b]}{[a b]},
\end{equation}
we can rewrite
\begin{equation}
  \operatorname{res}_{q_3^2 = q_4^2 = 0} \frac {d^4 q_1}{(q_1^2)^2 q_3^2 q_4^2} =
  \frac {d [1 \tilde{\lambda}] d [\tilde{\lambda} 4]}{-\langle 1 4\rangle^2 [1 4] [1 \tilde{\lambda}]^2 [\tilde{\lambda} 4]}.
\end{equation}
Now, taking the final residue in $q_1^2 = 0$ amounts to taking the residue in the variable $[1 \tilde{\lambda}]$ (because $q_1^2 = \langle 1 4\rangle [1 \tilde{\lambda}]$).  Hence, if there are no further terms that can produce a first order pole, the residue vanishes. Of course, introducing numerator factors can make it so that the residue taken here becomes nonzero. A similar calculation can be done on the second branch of the solution.

If the propagator in $q_1$ is \emph{not} squared, then taking the last two residues we find
\begin{equation}
  \operatorname{res}_{[\tilde{\lambda} 1] = [\tilde{\lambda} 4] = 0} \frac {d [\tilde{\lambda} 1] d [\tilde{\lambda} 4]}{-\langle 1 4\rangle [1 4] [\tilde{\lambda} 1] [\tilde{\lambda} 4]} =
  -\frac 1 {\langle 1 4\rangle [1 4]} =
  -\frac 1 {s_{2 3}}.
\end{equation}
This calculation yields the same answer (up to a permutation) as the similar calculation above for the case of a
non-squared propagator.

The calculation of the residue around $q_1^2 = q_3^2 = q_4^2 = 0$ for the case where there is no squared propagator can be done in a single step.  If we take one branch of the solution to $q_3^2 = q_4^2 = 0$ given by $q_3 = \lambda_4 (\tilde{\lambda} - \tilde{\lambda}_4)$ and $q_4 = \lambda_4 \tilde{\lambda}$ and we impose $q_1^2 = 0$ we find that we also need to impose $[1 \tilde{\lambda}] = 0$ so we have $\tilde{\lambda} = z \tilde{\lambda}_1$ for some complex number $z$.

We find
\begin{equation}
  \operatorname{res}_{q_1^2 = q_3^2 = q_4^2 = 0} \frac {d^4 q_1}{q_1^2 q_3^2 q_4^2} =
  \frac {d q_1^{2 \dot{2}}}{\frac {\partial (q_1^2, q_3^2, q_4^2)}{\partial (q_1^{1 \dot{1}}, q_1^{1 \dot{2}}, q_1^{2 \dot{1}})}} =
  \frac {d q_1^{2 \dot{2}}}{\det \begin{pmatrix}
    q_1^{2 \dot{2}} & q_3^{2 \dot{2}} & q_4^{2 \dot{2}} \\
    -q_1^{2 \dot{1}} & -q_3^{2 \dot{1}} & -q_4^{2 \dot{1}} \\
    -q_1^{1 \dot{2}} & -q_3^{1 \dot{2}} & -q_4^{1 \dot{2}}
  \end{pmatrix}} =
  \frac {d z}{-\langle 1 4\rangle [1 4] z},
\end{equation}
where $z$ is defined by the solution to the on-shell equations $\tilde{\lambda} = z \tilde{\lambda}_1$ above.  We have used $d q_1^{2 \dot{2}} = d q_4^{2 \dot{2}} = \lambda_4^2 \tilde{\lambda}_1^{\dot{2}} d z$ and the value of the determinant $-\langle 1 4\rangle \lambda_4^2 \tilde{\lambda}_1^{\dot{2}} [\tilde{\lambda} 4]$.

We can take another residue around $z = 0$ to find $-\frac 1 {s_{23}}$, which is the same as the result computed above by a different approach.  Importantly, the result is Lorentz-invariant.

Let us emphasize a few surprising aspects of the discussion above.
First, note that squaring a propagator, which one may fear makes the IR singularities worse, actually makes them better!  Indeed, explicit integrated results reveal that in an expansion around four dimensions the three one-mass triangles with a square propagator start at $\epsilon^{-1}$ instead of $\epsilon^{-2}$, while the triangle integrals with non-squared propagators start at $\epsilon^{-2}$.
Second, note that adding a numerator, which one may hope can only make things better (by canceling potential singularities), actually makes IR singularities worse (of order $\epsilon^{-2}$ instead of $\epsilon^{-1}$).

Both of these facts can be explained by how the residues behave.  Indeed, as previously mentioned, while a form $\frac {d z}{z^2}$ would appear more divergent than $\frac {d z}{z}$, it actually has zero residue.  However, it can contribute to the residue if it appears in a combination $\frac {d z}{z^2} f(z)$ (the analog of adding a numerator).

We also note that the triangle integrals with a squared propagator behave worse in an expansion around $D = 6$ than in four dimensions.  That is, increasing the dimension turns the divergence from order $\epsilon^{-1}$ to $\epsilon^{-2}$, while the opposite happens for the one-mass triangles with no squared propagators.  Even more surprisingly, all these integrals are \emph{finite} in an expansion around $D = 5$ and the box with a squared propagator \emph{vanishes} when $D \to 5$! 
The analytic results for the 5D boxes are given by
\begin{align}
   & I^{\text{(5D)}}(1,1,1,1)=-\frac{2 i \pi^4 \log \left(\frac{\sqrt{-s_{12}}+\sqrt{-s_{23}}+\sqrt{-s_{12}-s_{23}}}{\sqrt{-s_{12}}+\sqrt{-s_{23}}-\sqrt{-s_{12}-s_{23}}}\right)}{\sqrt{-s_{12} s_{23} (s_{12}+s_{23})}}+\mathcal{O}(\epsilon)\,,
       \\
 &\nonumber I^{\text{(5D)}}(2,1,1,1)\\&=-\epsilon\frac{4 i \pi^4 s_{12} \sqrt{-s_{23}} (s_{12}+s_{23})}{s_{12}^2 s_{23}^2 (s_{12}+s_{23})}\\&\nonumber-\epsilon\frac{4 i \pi^4 s_{23} \sqrt{-s_{12} s_{23} (s_{12}+s_{23})} \log \left(\frac{\sqrt{-s_{12}}+\sqrt{-s_{23}}+\sqrt{-s_{12}-s_{23}}}{\sqrt{-s_{12}}+\sqrt{-s_{23}}-\sqrt{-s_{12}-s_{23}}}\right)}{s_{12}^2 s_{23}^2 (s_{12}+s_{23})}+\mathcal{O}(\epsilon^2)\,.
\end{align}
Note that we do not include factors of $\pi$ in the definition of integrals (going against a widespread convention since, in particular, it is not convenient in odd dimensions).  Hence, the expressions we obtain for integrals always have weight smaller or equal to the number of integrations.

The integrals above contain $\sqrt{-s_{12} - s_{23}}$, but this singularity actually cancels between the logarithm and the prefactor (each produces a minus sign).  This singularity does appear after a monodromy around $s_{12} = 0$ or $s_{23} = 0$.  For $I^{\text{(5D)}}(2, 1, 1, 1)$ also the pole at $s_{12} + s_{23} = 0$ cancels.

We emphasize that even though all calculations we performed so far are at one-loop order, its applicability is wider (see sec.~\ref{sec:two-loop-example} for some examples).  Indeed, one can put in the four-point vertex any (including non-planar) integrand with four-point external kinematics.  Of course, to establish IR-convergence one needs in general to impose the cancellation of potentially many more permanent pinches.
\subsection{Triangle integrals in arbitrary integer dimensions}
\label{sec:residue-other-dims}

Consider again the triangle pinch on the triangle topology with no dotted propagators, but now in $d$ dimensions (where we take $d$ to be an integer). We express the computation in dual coordinates $x_i$, defined such that $p_i=x_i-x_{i-1}$, and we indicate the loop integration by $x_0$. In these coordinates, the soft pinch $q_1=0$ corresponds to $x_1=x_0$. We want to compute
\begin{equation}
  \operatorname{res} \frac {d^d x_0}{x_{01}^2 x_{02}^2 x_{04}^2},
\end{equation}
where $x_{12}^2 = x_{14}^2 = 0$ but $x_{24}^2 \neq 0$.

We will use a coordinate system with
\begin{gather}
    x_1 = (0, 0, \vec{0}), \qquad
    x_2 = (-x_{12}^-, 0, \vec{0}), \\
    x_4 = (0, -x_{14}^+, \vec{0}), \qquad
    x_0 = (\alpha, \beta, z \vec{y})
\end{gather}
where the first two components are light-cone coordinates $+$ and $-$ and the last $d-2$ coordinates form a (transverse) vector.  We take $\|y\| = 1$.

The rationale for this parametrization is that when $\alpha = \beta = z = 0$ we recover the pinch kinematics $x_0 = x_1$ and we do this by imposing three conditions (which is the same as the number of propagators in the permanent pinch Landau diagram).  When $z = 0$ the $(d-3)$-dimensional sphere parameterized by $y$ will decouple and the integral in $y$ can be done trivially.

In these coordinates we have
\begin{gather}
    x_{01}^2 = \alpha \beta - z^2, \\
    x_{02}^2 = (\alpha + x_{14}^+) \beta - z^2 \\
    x_{04}^2 = \alpha (\beta + x_{12}^-) - z^2, \\
    d^D x_0 = d \alpha d \beta z^{D - 3} d z d^{D - 3} \Omega(\vec{y}).
\end{gather}

The residue reads
\begin{equation}
    \operatorname{res} \frac {d^d x_0}{x_{01}^2 x_{02}^2 x_{04}^2} =
    \frac{d \alpha d \beta z^{d - 3} d z d^{d - 3} \Omega(\vec{y})}{\frac{\partial (x_{01}^2, x_{02}^2, x_{04}^2)}{\partial (\alpha, \beta, z)} d \alpha d \beta d z} =
    \frac 1 {2 x_{12}^- x_{14}^+} z^{d - 4} d^{d - 3} \Omega(\vec{y}).
\end{equation}
As $x_{12}^- x_{14}^+ = -(x_2 - x_4)^2=-s_{12}$, we see that for $d = 4$ we obtain the same result as computed in eq.\eqref{eq:triangle-trianglepinch} of the previous section, while for $d > 4$ the residue vanishes (since we evaluate the answer at $\alpha = \beta = z = 0$).  In particular, the answer is Lorentz-invariant, even though the intermediate steps of the calculation explicitly break the Lorentz symmetry.

\begin{figure}
     \centering
     \begin{subfigure}[m]{0.3\textwidth}
         \centering
         \includegraphics[width=\textwidth]{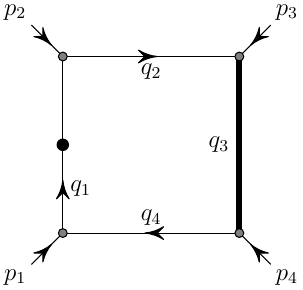}
       \end{subfigure}
     \hfil
     \begin{subfigure}[m]{0.3\textwidth}
         \centering
         \includegraphics[width=\textwidth]{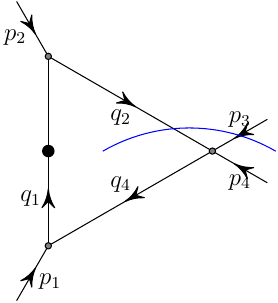}
     \end{subfigure}
     \caption{A box with a squared propagator and the opposite propagator massive and a triangle with a squared propagator and a non-trivial numerator factor. The blue arc between two regions in the left diagram corresponds to $q_3$ appearing as its numerator.}
     \label{fig:squared-massive-box-triangle-num}
\end{figure}
\subsection{Box integrals}
As an example of nontrivial cancellation of IR singularities, consider the box integral with a squared propagator and a mass on the opposite propagator (see fig.~\ref{fig:squared-massive-box-triangle-num}).  We will show that for the triangle permanent pinch we can cancel the leading divergence ($\epsilon^{-2}$ in an expansion around $D = 4$) by the triangle in fig.~\ref{fig:squared-massive-box-triangle-num}.

Indeed, by the same methods described above, one can compute the residue in the triangle kinematics to find that when replacing the propagator $\frac 1 {q_3^2}$ by the massive propagator $\frac 1 {q_3^2 - m^2}$ the residue becomes
\begin{equation}
  \frac{m^2 s_{1 2} + 2 m^2 s_{2 3} + s_{1 2} s_{2 3}}{s_{1 2}^2 (m^2-s_{2 3})^3}.
\end{equation}
When $m = 0$ this reproduces $-\frac 1 {s_{1 2} s_{2 3}^2}$ computed above.  The residue of the triangle is the same as computed above.  Using the notation introduced in sec.~\ref{sec:simple-examples},
our derivation shows that the difference
\begin{equation}
  I_m(2, 1, 1, 1) - \frac {m^2 s_{1 2} + 2 m^2 s_{2 3} + s_{1 2} s_{2 3}}{(s_{1 2} + 2 s_{2 3}) (s_{2 3} - m^2)^3} I_m(2, 1, -1, 1)
\end{equation}
has no triangle permanent pinches.  Its expansion around $D = 4$ starts at order $\epsilon^{-1}$, so canceling the triangle permanent pinches canceled the leading divergence in the $\epsilon$ expansion.

There are still two bubble permanent pinches, so let us analyze them next.  We expect that canceling these bubble singularities will cancel the divergent terms of order $\epsilon^{-1}$ (except for the contribution coming from the squared tadpole singularity, which is treated in sec.~\ref{sec:squared-tadpoles}).

For this topology, there is a bubble permanent pinch enforcing collinearity of $q_1$, $q_4$ with $p_1$, and another of $q_1$, $q_2$ with $p_2$; since these are related by symmetry, we only need to analyze one.  For the permanent pinch in the $p_1$ channel, we want to take residues of type
\begin{equation}
  \operatorname{res}_{q_1^2 = q_4^2 = 0} \frac {d^4 q_4}{(q_1^2)^2 q_4^2} \Bigl(\cdots\Bigr),
\end{equation}
where the quantity in brackets will not be important for now.  First, we take the residue in $q_4^2$.  We have
\begin{equation}
  \operatorname{res}_{q_4^2 = 0} \frac {d^4 q_4}{q_4^2} = \left.\frac {d q_4^{1 \dot{2}} d q_4^{2 \dot{1}} d q_4^{2 \dot{2}}}{q_4^{2 \dot{2}}}\right\rvert_{q_4^2 = 0}.
\end{equation}
We can solve the massless constraint by setting $q_4^{\alpha \dot{\alpha}} = \lambda^\alpha \tilde{\lambda}^{\dot{\alpha}}$ which yields
\begin{equation}
  \operatorname{res}_{q_4^2 = 0} \frac {d^4 q_4}{q_4^2} =
  d \lambda^1 d \lambda^2 (\tilde{\lambda}^{\dot{1}} d \tilde{\lambda}^{\dot{2}} - \tilde{\lambda}^{\dot{2}} d \tilde{\lambda}^{\dot{1}}) -
  (\lambda^1 d \lambda^2 - \lambda^2 d \lambda^1) d \tilde{\lambda}^{\dot{1}} d \tilde{\lambda}^{\dot{2}}.
\end{equation}
Next, we want to take the residue in $q_1^2$ which is more involved since this is a second order pole.  We need to compute\footnote{The integration measure can be obtained by contracting into the canonical measure $d^2 \lambda d^2 \tilde{\lambda}$ a vector field $\lambda \cdot \partial_\lambda - \tilde{\lambda} \cdot \partial_{\tilde{\lambda}}$, which encodes the transformations of $\lambda$ and $\tilde{\lambda}$ with opposite weights.  This is analogous to constructing the measure on $\mathbb{P}^n$ by contracting the measure on a $\mathbb{C}^{n + 1}$ with the Euler vector field.}
\begin{equation}
  \operatorname{res}_{q_1^2 = 0} \frac {d^2 \lambda [\tilde{\lambda} d \tilde{\lambda}] -
  \langle \lambda d \lambda\rangle d^2 \tilde{\lambda}}{\langle \lambda \lambda_1\rangle^2 [\tilde{\lambda} \tilde{\lambda}_1]^2} \Bigl(\cdots\Bigr).
\end{equation}

We rewrite this slightly, via two identities (one of which we have seen before)
\begin{gather}
  \frac {d^2 \tilde{\lambda}}{[\tilde{\lambda} \tilde{\lambda}_1]^2} = \frac {d [\tilde{\lambda} \tilde{\lambda}_1]}{[\tilde{\lambda} \tilde{\lambda}_1]^2} \frac {d [\tilde{\lambda} \tilde{\lambda}_2]}{[1 2]}, \\
  \frac {\langle \lambda d \lambda\rangle}{\langle \lambda \lambda_1\rangle^2} = \frac {\langle \lambda_1 \lambda\rangle d \langle \lambda_2 \lambda\rangle - \langle \lambda_2 \lambda\rangle d \langle \lambda_1 \lambda\rangle}{\langle 1 2\rangle \langle \lambda \lambda_1\rangle^2},
\end{gather}
where $\lambda_2$, $\tilde{\lambda}_2$ can be thought of as some auxiliary spinors; the dependence on those will cancel in the final answer.  We now take residues in the variables $\langle \lambda \lambda_1\rangle$ and $[\tilde{\lambda} \tilde{\lambda}_1]$.  Taking these residues makes $\lambda \tilde{\lambda}$ collinear with $\lambda_1 \tilde{\lambda}_1$, which is the permanent pinch kinematics.  At the location of the permanent pinch, we have $\lambda \tilde{\lambda} = z \lambda_1 \tilde{\lambda}_1$.

We take the new variables to be
\begin{gather}
  v = \frac {\langle \lambda \lambda_2\rangle}{\langle 1 2\rangle}, \qquad
  \tilde{v} = \frac {[\tilde{\lambda} \tilde{\lambda}_2]}{[1 2]}, \\
  u = \langle \lambda \lambda_1\rangle, \qquad
  \tilde{u} = [\tilde{\lambda} \tilde{\lambda}_1].
\end{gather}
At the location of the permanent pinch we have $v \tilde{v} = z$.  Indeed, even though it looks like the residue should depend on $v$ and $\tilde{v}$, it will turn out that it actually only depends on their product (and therefore on $z$).

We find
\begin{gather}
  \frac {d^2 \tilde{\lambda}}{[\tilde{\lambda} \tilde{\lambda}_1]^2} =
  \frac {d \tilde{u}}{\tilde{u}^2} d \tilde{v}, \qquad
  \frac {\langle \lambda d \lambda\rangle}{\langle \lambda \lambda_1\rangle^2} =
  \frac {u d v - v d u}{u^2}, \\
  \frac {d^2 \lambda}{\langle \lambda \lambda_1\rangle^2} = \frac {d u}{u^2} d v, \qquad
  \frac {[\tilde{\lambda} d \tilde{\lambda}]}{[\tilde{\lambda} \tilde{\lambda}_1]^2} = \frac {\tilde{u} d \tilde{v} - \tilde{v} d \tilde{u}}{\tilde{u}^2}.
\end{gather}

In the new variables, the residue becomes
\begin{equation}
  \operatorname{res}_{u = \tilde{u} = 0} \frac {d u d \tilde{u}}{u^2 \tilde{u}^2} d (v \tilde{v}) \Bigl(\cdots\Bigr).
\end{equation}

In order to take the residue, we need to express brackets $\langle \lambda \lambda_i\rangle$ and $[\tilde{\lambda} \tilde{\lambda}_i]$ in terms of the variables $u, \tilde{u}, v, \tilde{v}$.  We have
\begin{gather}
  \langle \lambda \lambda_i\rangle = -u \frac {\langle 2 i\rangle}{\langle 1 2\rangle} + \langle 1 i\rangle v, \\
  [\tilde{\lambda} \tilde{\lambda}_i] = -\tilde{u} \frac {[2 i]}{[1 2]} + [1 i] \tilde{v}.
\end{gather}

Hence,
\begin{gather}
  q_2^2 = (q_4 + p_1 + p_2)^2 = \langle \lambda \lambda_1\rangle [\tilde{\lambda} \tilde{\lambda}_1] + \langle \lambda \lambda_2\rangle [\tilde{\lambda} \tilde{\lambda}_2] + s_{1 2} =
  u \tilde{u} + s_{1 2} (1 + v \tilde{v}), \\
  q_3^2 = (q_4 - p_4)^2 = -\Bigl(-u \frac {\langle 24\rangle}{\langle 12\rangle} + \langle 14\rangle v\Bigr) \Bigl(-\tilde{u} \frac {[2 4]}{[1 2]} + [1 4] \tilde{v}\Bigr).
\end{gather}

We want to compute a residue
\begin{equation}
  \operatorname{res}_{u = \tilde{u} = 0} \frac {d u d \tilde{u} d (v \tilde{v}) + d (u \tilde{u}) d v d \tilde{v}}{u^2 \tilde{u}^2} f(u, \tilde{u}, v, \tilde{v}),
\end{equation}
where the function $f$ has the homogeneity property $f(u, \tilde{u}, v, \tilde{v}) = f(t u, t^{-1} \tilde{u}, t v, t^{-1} \tilde{v})$ for all non-vanishing $t$.  By taking $t = \tilde{v}$, we find that $f$ is a function of three variables $f(u, \tilde{u}, v, \tilde{v}) = f(u \tilde{v}, \tilde{u} \tilde{v}^{-1}, v \tilde{v}, 1)$.

We introduce the notation $\rho = u \tilde{v}$, $\sigma = \tilde{u} \tilde{v}^{-1}$ and $z = v \tilde{v}$.  In terms of these new variables, the differential form in the numerator can be rewritten as
\begin{equation}
  d u d \tilde{u} d (v \tilde{v}) + d (u \tilde{u}) d v d \tilde{v} =
  d \rho d \sigma d z.
\end{equation}
Hence, we need to compute the residue
\begin{equation}
  \operatorname{res}_{\rho = \sigma = 0} \frac {d \rho d \sigma d z}{\rho^2 \sigma^2} F(\rho, \sigma, z),
\end{equation}
where $F(\rho, \sigma, z) = f(\rho, \sigma, z, 1)$.

Under a change of reference spinors, we have
\begin{gather}
  v \to v' = v + \xi u, \qquad
  \tilde{v} \to \tilde{v}' = \tilde{v} + \tilde{\xi} \tilde{u},
\end{gather}
which implies a change of variables
\begin{equation}
  (\rho, \sigma, z) \to (\rho', \sigma', z') =
  (\rho (1 + \tilde{\xi} \sigma),
  \sigma (1 + \tilde{\xi} \sigma)^{-1},
  z + \xi \rho + \tilde{\xi} \sigma z + \xi \tilde{\xi} \rho \sigma).
\end{equation}
We have that the Jacobian is unity so
\begin{equation}
  d \rho' d \sigma' d z' = d \rho d \sigma d z
\end{equation}
and also ${\rho'}^2 {\sigma'}^2 = \rho^2 \sigma^2$. Hence,
\begin{equation}
  \frac {d \rho d \sigma d z}{\rho^2 \sigma^2} F(\rho', \sigma', z') =
  \frac {d \rho' d \sigma' d z'}{{\rho'}^2 {\sigma'}^2} F(\rho', \sigma', z').
\end{equation}
Taking the residue in the $(\rho, \sigma)$ variables we get
\begin{equation}
  d z \partial_\rho \partial_\sigma F(0, 0, z),
\end{equation}
and similarly for $z'$.  To compare, we recall that when $\rho = \sigma = 0$ then $z' = z$.

Here it is essential to remember that the result of taking the residue is a one-form.  If we remove the $d z$ factor then the resulting quantity does \emph{not} transform nicely under change of reference spinors.

Applying the formulas above we obtain the residue one-form for the box in fig.~\ref{fig:squared-massive-box-triangle-num}
\begin{equation}
\begin{aligned}
  &\frac{2 m^2 s_{2 3} (s_{1 2}+s_{2 3}) d z}
  {s_{1 2}^2 \left(m^2-s_{2 3}\right) \left(z s_{2 3}+m^2\right)^3}\\&+
  \frac{s_{2 3} \left(m^2 s_{1 2}+2 m^2 s_{2 3}+s_{1 2} s_{2 3}\right) d z}
  {s_{1 2}^2 \left(m^2-s_{2 3}\right)^3 \left(z s_{2 3}+m^2\right)}-
  \frac{(m^2 s_{1 2}+2 m^2 s_{2 3}+s_{1 2} s_{2 3}) d z}
  {(z+1) s_{1 2}^2 \left(m^2-s_{2 3}\right)^3}\\&+
  \frac{d z}{(z+1)^2 s_{1 2}^2 \left(m^2-s_{2 3}\right)} +
  \frac{s_{2 3} (s_{1 2}+s_{2 3}) \left(s_{2 3}+m^2\right) d z}
  {s_{1 2}^2 \left(m^2-s_{2 3}\right)^2 \left(z s_{2 3}+m^2\right)^2}.
\end{aligned}
\end{equation}
Some other residues in the bubble $p_1$ permanent pinch kinematics are
\begin{align}
  \operatorname{res}_{q_1^2 = q_4^2 = 0} I_m(2, 1, -1, 1) &= d z\Bigl(\frac{m^2-s_{2 3}}{(z+1)^2 s_{1 2}^2}+\frac{s_{1 2}+2 s_{2 3}}{(z+1) s_{1 2}^2}\Bigr), \\
  \operatorname{res}_{q_1^2 = q_4^2 = 0} I_m(2, 0, 1, 1) &= \frac{(s_{1 2}+s_{2 3}) d z}{s_{1 2} \left(z s_{2 3}+m^2\right)^2}-\frac{2 m^2 (s_{1 2}+s_{2 3}) d z}{s_{1 2} \left(z s_{2 3}+m^2\right)^3}, \\
  \operatorname{res}_{q_1^2 = q_4^2 = 0} I_m(1, 1, 1, 1) &= \frac{s_{2 3} d z}{s_{1 2} \left(m^2-s_{2 3}\right) \left(z s_{2 3}+m^2\right)}-\frac{d z}{(z+1) s_{1 2} \left(m^2-s_{2 3}\right)}, \\
  \operatorname{res}_{q_1^2 = q_4^2 = 0} I_m(1, 1, -1, 1) &= -\frac{(m^2-s_{2 3}) d z}{(z+1) s_{1 2}}-\frac{s_{2 3} d z}{s_{1 2}}, \\
  \operatorname{res}_{q_1^2 = q_4^2 = 0} I_m(1, 0, 1, 1) &= -\frac{d z}{z s_{2 3}+m^2}, \\
  \operatorname{res}_{q_1^2 = q_4^2 = 0} I_m(1, 1, 0, 1) &= \frac{d z}{(z+1) s_{1 2}}.
\end{align}

We record here some residues in the triangle permanent pinch
\begin{align}
  \operatorname{res}_{q_1^2 = q_2^2 = q_4^2 = 0} I_m(2, 1, 1, 1) &=
  \frac {m^2 s_{1 2} + 2 m^2 s_{2 3} + s_{1 2} s_{2 3}}{s_{1 2}^2 (m^2 - s_{2 3})^3}, \\
  \operatorname{res}_{q_1^2 = q_2^2 = q_4^2 = 0} I_m(2, 1, -1, 1) &= -\frac {s_{1 2} + 2 s_{2 3}}{s_{1 2}^2}, \\
  \operatorname{res}_{q_1^2 = q_2^2 = q_4^2 = 0} I_m(1, 1, 0, 1) &= -\frac 1 {s_{1 2}}, \\
  \operatorname{res}_{q_1^2 = q_2^2 = q_4^2 = 0} I_m(1, 1, 1, 1) &= \frac 1 {s_{1 2} (m^2 - s_{2 3})}, \\
  \operatorname{res}_{q_1^2 = q_2^2 = q_4^2 = 0} I_m(1, 1, -1, 1) &= \frac {m^2 - s_{2 3}}{s_{1 2}}.
\end{align}

The bubble pinches impose several conditions since one has to match a function of one variable, so after partial fractioning one needs to independently match the coefficients of $(1 + z)^{-1}$, $(1 + z)^{-2}$, $(m^2 + z s_{2 3})^{-1}$, $(m^2 + z s_{2 3})^{-2}$ and $(m^2 + z s_{2 3})^{-3}$.

Another observation to make here is that taking the residue at $z = -1$ for the bubble residue we obtain the triangle residues (up to a sign). In keeping with the procedure stated in sec.~\ref{sec:intro}, this again supports the claim that the soft configurations which are part of a family of collinear configurations do not need to be considered separately, except for cases in which relevant tadpole-like pinches appear.

A solution to the vanishing residues in all triangle and bubble permanent pinch configurations is:
\begin{equation}
\begin{aligned}
  &I_m(2,1,1,1) + c_{1}I_m(2,1,-1,1) - c_{2,-2} I_m(2,1,0,1) - 
   c_{3,-3} (I_m(1,0,3,1) + I_m(1,1,3,0))\\& - 
   c_{3,-2} (I_m(1,0,2,1) + I_m(1,1,2,0)) - 
   c_{3,-1} (I_m(1,0,1,1) + I_m(1,1,1,0)),
\end{aligned}
\label{eq:box-finite-comb}
\end{equation}
 where
\begin{align}
  c_1 &= \frac{m^2 (-s_{1 2})-2 m^2 s_{2 3}-s_{1 2} s_{2 3}}{(s_{1 2}+2 s_{2 3})
    \left(s_{2 3}-m^2\right)^3}, \\
  c_{2,-2} &= -\frac{2 \left(m^2 s_{1 2}+2 m^2 s_{2 3}-s_{2 3}^2\right)}{(s_{1 2}+2 s_{2 3}) \left(m^2-s_{2 3}\right)^2}, \\
  c_{3,-3} &= -\frac{2 m^2 s_{2 3} (s_{1 2}+s_{2 3})}{s_{1 2}^2 \left(m^2-s_{2 3}\right)}, \\
  c_{3,-2} &= \frac{s_{2 3} (s_{1 2}+s_{2 3}) \left(s_{2 3}+m^2\right)}{s_{1 2}^2
    \left(m^2-s_{2 3}\right)^2}, \\
  c_{3,-1} &= -\frac{s_{2 3} \left(m^2 s_{1 2}+2 m^2 s_{2 3}+s_{1 2} s_{2 3}\right)}{s_{1 2}^2 \left(m^2-s_{2 3}\right)^3}.
\end{align}

However, this still has a squared tadpole permanent pinch singularity, so it is still IR-divergent. For usual tadpoles in high enough dimension, these permanent pinches yield zero residues, and are thus harmless. In fact, we didn't need to consider this type of pinch in any previous example. In cases such as the present one, in which this is instead necessary, the computation of the corresponding residues turns out to be surprisingly involved. We treat it in detail in sec.~\ref{sec:squared-tadpoles}.  For now it is sufficient to note that, in dimensional regularization, one can see directly from integrated values that the final divergence of the integral combination in~\eqref{eq:box-finite-comb} can be canceled by a $t$-channel bubble with one squared and one massive propagator.  The coefficient of this integral is $\frac{2 s_{23}^2}{s_{12}^2 (s_{23} - m^2)^2}$.

We present the divergent parts of some integrals of this family in dimensional regularization in table~\ref{tab:integrals}.  The entry marked with $\ast$ reads
\begin{equation}
\begin{aligned}
\label{eq:I2111}
  &\frac{\left(m^2 s_{1 2}+2 m^2 s_{2 3}+s_{1 2} s_{2 3}\right) \log \left(-\frac{s_{1 2}}{\mu^2}\right)}{s_{1 2}^2 \left(m^2-s_{2 3}\right)^3} \\&
  + \frac{2\left(m^2 s_{1 2}+2 m^2 s_{2 3}+s_{1 2} s_{2 3}\right) \log
    \left(\frac{m^2-s_{2 3}}{m^2}\right)}{s_{1 2}^2 \left(m^2-s_{2 3}\right)^3} 
  -\frac{2 \left(-m^4+2 m^2 s_{2 3}+3 s_{1 2} s_{2 3}+s_{2 3}^2\right)}{s_{1 2}^2 \left(m^2-s_{2 3}\right)^3}.
  \end{aligned}
\end{equation}

\begin{table}
  \centering
  \begin{tabular}{l|c|c}
    integral & $\epsilon^{-2}$ & $\epsilon^{-1}$ \\
    \hline
    $I_m(2, 1, 1, 1)$ & $-\frac {m^2 s_{1 2} + 2 m^2 s_{2 3} + s_{1 2} s_{2 3}}{s_{1 2}^2 (m^2 - s_{2 3})^3}$ & $\ast$ \\
    \hline
    $I_m(2, 1, -1, 1)$ & $\frac {s_{1 2} + 2 s_{2 3}}{s_{1 2}^2}$ & $s_{1 2}^{-2} \Bigl(2 (m^2 - s_{2 3}) + (s_{1 2} + 2 s_{2 3}) \log \frac {\mu^2}{-s_{1 2}}\Bigr)$ \\
    \hline
    $I_m(1, 0, 3, 1)$ & $0$ & $\frac {2 m^2 - s_{2 3}}{2 m^4 (m^2 - s_{2 3})^2}$ \\
    \hline
    $I_m(1, 1, 3, 0)$ & $0$ & $\frac {2 m^2 - s_{2 3}}{2 m^4 (m^2 - s_{2 3})^2}$ \\
    \hline
    $I_m(1, 0, 2, 1)$ & $0$ & $-\frac 1 {m^2 (m^2 - s_{2 3})}$ \\
    \hline
    $I_m(1, 1, 2, 0)$ & $0$ & $-\frac 1 {m^2 (m^2 - s_{2 3})}$ \\
    \hline
    $I_m(1, 0, 1, 1)$ & $0$ & $-s_{2 3}^{-1} \log (1 - \frac {s_{2 3}}{m^2})$ \\
    \hline
    $I_m(1, 1, 1, 0)$ & $0$ & $-s_{2 3}^{-1} \log (1 - \frac {s_{2 3}}{m^2})$ \\
    \hline
    $I_m(2, 1, 0, 1)$ & $0$ & $-\frac 2 {s_{1 2}^2}$ \\
    \hline
  \end{tabular}
  \caption{$\epsilon$ expansion for a few integrals. The expression for the entry marked with $\ast$ can be found in eq.~\eqref{eq:I2111}. These values are obtained using the normalization prescription of~\cite{Ellis:2007qk}, in particular including factors of $\pi$ in the denominator in the integral definition.}
  \label{tab:integrals}
\end{table}

\subsection{Cohomology class of residues and simple representatives}
\label{sec:cohomology}
As discussed in appendix~\ref{sec:higher-residues}, only the cohomology class of the residue is intrinsically well defined.  In the calculations of the bubble permanent pinch, the residues are elements of a relative cohomology of $\mathbb{C}$ minus the points where the differential forms have poles. These occur at $z = -1$ and $z = -\frac{m^2}{s_{23}}$.  The two singularities play different roles: while $z = -\frac{m^2}{s_{23}}$ can be easily crossed by an appropriate choice of contour, its residue manifesting in the discontinuity of the logarithm appearing in the integrated answer, the pole at $z = -1$ instead corresponds to the additional singularity identified by the triangle permanent pinch.

We will pair cohomology classes of the residues with homology classes of the permanent pinch.  It can be shown (by using techniques described in refs.~\cite{AIHPA_1967__6_2_89_0, Hannesdottir:2022xki}) that, for Feynman propagators, the bubble permanent pinch traps the contour for $z \in [-1, 0]$.  From this perspective, it is natural to take the residues to be elements of a \emph{relative} cohomology, that is consider differential one-forms modulo total derivatives of functions which vanish at the boundaries $z = 0$ and $z = -1$. In the following, we will show that the bubble permanent pinches can \emph{also} be canceled by other integrals (with lower powers of propagators) whose bubble residue is cohomologous to the one computed above.  Moreover, we will show that the scheme independent kinematic dependence of the leading divergent part can be computed by simply integrating the residue cohomology class along a contour from $0$ to $-1$.  The ambiguity in the choice of contour reflects the multivalued nature of the leading divergent parts.

Consider the relative cohomology group $H^1(\mathbb{C} \setminus \{a\}; \{b, c\})$.  This is relevant for us in the study of bubble pinches where the variable parameterizing $\mathbb{C}$ is the momentum fraction and $a = -\frac {m^2}{s_{2 3}}$, $b = 0$ and $c = -1$ (in our choice of parameterization).

When partial-fractioning the residue form we obtain higher powers such as $\frac {d z}{(z - a)^k}$.  They can be reduced to a basis of the relative cohomology as follows.

We pick the cohomology basis $[\omega_0] = \left[\frac {d z}{c - b}\right]$ and $[\omega_1] = \left[\frac {d z}{z - a}\right]$ and an integral homology basis consisting of $\gamma_0$, a small loop around $z = a$, and $\gamma_1$, a path from $b$ to $c$.  We have
\begin{equation}
  \begin{pmatrix}
    \int_{\gamma_0} \omega_0 & \int_{\gamma_0} \omega_1 \\
    \int_{\gamma_1} \omega_0 & \int_{\gamma_1} \omega_1
  \end{pmatrix} =
  \begin{pmatrix}
    0 & 2 \pi i \\
   1& \log \frac {c - a}{b - a} 
  \end{pmatrix}
\end{equation}

If $k \neq 1$ we have that
\begin{gather}
  \int_{\gamma_0} \frac {d z}{(z - a)^k} = 0, \qquad
  \int_{\gamma_1} \frac {d z}{(z - a)^k} = \frac 1 {-k + 1} \Bigl((c - a)^{-k + 1} - (b - a)^{-k + 1}\Bigr),
\end{gather}
which implies (using the fact that the pairing between homology and cohomology is perfect) that
\begin{equation}
  \label{eq:relative-cohomology-reduction}
  \left[\frac {d z}{(z - a)^k}\right] =
  \frac {(c - a)^{-k + 1} - (b - a)^{-k + 1}}{(-k + 1) (c - b)} [d z].
\end{equation}
Here, the equality should be interpreted in the relative cohomology.  In this case, this means that the two differential forms differ not just by a total derivative, but by the total derivative of a function that vanishes at $z = b$ and $z = c$.  We note that in absolute (non-relative) cohomology, we would have instead $\frac {d z}{(z - a)^k} = d \Bigl(\frac {(z - a)^{-k + 1}}{-k + 1}\Bigr)$, so that the form would be trivial.

Using this, we find
\begin{multline}
  \operatorname{Res}_{q_1^2 = q_4^2 = 0} I_m(2, 1, 1, 1)  \\=
  \frac 1 {s_{1 2}^2 (m^2 - s_{2 3})} \left[\frac {d z}{(1 + z)^2}\right] -
  \frac {m^2 s_{1 2} + 2 m^2 s_{2 3} + s_{1 2} s_{2 3}}{s_{1 2}^2 (m^2 - s_{2 3})^3} \left[\frac {d z}{1 + z}\right] \\+
  \frac {3 s_{2 3} (s_{1 2} + s_{2 3})}{s_{1 2}^2 (m^2 - s_{2 3})^3} [d z] +
  \frac {s_{2 3} (m^2 s_{1 2} + 2 m^2 s_{2 3} + s_{1 2} s_{2 3})}{s_{1 2}^2 (m^2 - s_{2 3})^3} \left[\frac {d z}{m^2 + z s_{2 3}}\right],
\end{multline}
where we have used the notation ``$\operatorname{Res}$'' for the cohomology class of the residue, and we kept expressions that are singular at the boundary.  It is a theorem that only the cohomology class of the residue is unambiguously defined.

We present below the reduction to the relative cohomology basis of a few of the residues computed above.  We only list the ones for which the reduction differs significantly from the original writing:
\begin{gather}
  \operatorname{Res}_{q_1^2 = q_4^2 = 0} I_m(2, 1, -1, 1) = \frac{m^2-s_{2 3}}{s_{1 2}^2} \left[\frac {d z}{(z + 1)^2}\right] + \frac{s_{1 2}+2 s_{2 3}}{s_{1 2}^2} \left[\frac {d z}{z + 1}\right], \\
  \operatorname{Res}_{q_1^2 = q_4^2 = 0} I_m(2, 0, 1, 1) = -\frac{s_{1 2}+s_{2 3}}{s_{1 2} \left(m^2-s_{2 3}\right)^2} [d z], \\
  \operatorname{Res}_{q_1^2 = q_4^2 = 0} I_m(1, 1, 1, 1) = \frac{s_{23}}{s_{12} (m^2 - s_{23})} \Bigl[\frac{d z}{z s_{23} + m^2}\Bigr] - \frac{1}{s_{12} (m^2 - s_{23})} \Bigl[\frac{d z}{z + 1}\Bigr], \\
  \operatorname{Res}_{q_1^2 = q_4^2 = 0} I_m(1, 1, -1, 1) = -\frac{m^2 - s_{23}}{s_{12}} \Bigl[\frac{d z}{z + 1}\Bigr] - \frac{s_{23}}{s_{12}} [d z], \\
  \operatorname{Res}_{q_1^2 = q_4^2 = 0} I_m(1, 0, 1, 1) = -\Bigl[\frac{d z}{z s_{23} + m^2}\Bigr], \\
  \operatorname{Res}_{q_1^2 = q_4^2 = 0} I_m(1, 1, 0, 1) = \frac{1}{s_{12}} \Bigl[\frac{d z}{z + 1}\Bigr].
\end{gather}

We can cancel all the bubble poles in the linear combination
\begin{multline}
    I_m(2,1,1,1)-\frac{1}{(m^2-s_{23})^2}I_m(2,1,-1,1)+\frac{2(m^2s_{12}+2m^2s_{23}-s_{23}^2)}{{s_{12}}(m^2-s_{23})^3}I_m(1,1,0,1)+\\
    \frac{s_{23}(m^2s_{12}+2m^2s_{23}+s_{12}s_{23})}{s_{12}^2(m^2-s_{23})^3}(I_m(1,0,1,1)+I_m(1,1,1,0))+\\
    \frac{3s_{23}}{s_{12}(m^2-s_{23})}(I_m(2,0,1,1)+I_m(2,1,1,0)),
\end{multline}
where we have also subtracted the poles arising in the other bubble permanent pinch where the propagators $q_1^2$ and $q_2^2$ are on-shell. One can check that the triangle residue also vanishes for this linear combination.

By integrating the expressions analytically in dimensional regularization, we find that this linear combination has a divergence in an expansion around $D = 4$
\begin{equation}
    \frac 1 \epsilon \frac{4 s_{23}^2}{s_{12}^2 (s_{23}-m^2)^3},
\end{equation}
so we indeed have successfully removed the $\epsilon^{-2}$ divergence. This remaining contribution can only arise from the tadpole-like pinch of this topology, which is the topic of the following section.

\section{Tadpole-like soft singularities}
\label{sec:squared-tadpoles}
\subsection{Tadpole in a box with a massive propagator}
We now present the calculation of a squared tadpole permanent pinch for a box integral with one doubled propagator and one massive propagator.  In dual space notation, the form to integrate reads
\begin{equation}
  \phi = \frac {d^4 x_0}{(x_{0 1}^2)^2 x_{0 2}^2 (x_{0 3}^2 - m^2) x_{0 4}^2}.
\end{equation}

We can choose a parametrization of the box integral $x_0 = x_1 + z (1, \vec{y})$ (a similar parameterization was used in ref.~\cite{Vergu:2025mag}).  This corresponds to a blow-up of the origin for the light-cone of equation $x_{01}^2 = 0$.  Then, we have
\begin{gather}
  x_{01}^2 = z^2 (1 - \vec{y}^{\,2}), \\
  x_{02}^2 = 2 z x_{12} \cdot (1, \vec{y}) + z^2 (1 - \vec{y}^2), \\
  x_{04}^2 = 2 z x_{14} \cdot (1, \vec{y}) + z^2 (1 - \vec{y}^2), \\
  x_{03}^2 = x_{13}^2 + 2 z x_{13} \cdot (1, \vec{y}) + z^2 (1 - \vec{y}^2),
\end{gather}
where we have used $x_{12}^2 = x_{41}^2 = 0$.  Then,
\begin{multline}
  \omega = \frac {d^4 x_0}{x_{01}^2 x_{02}^2 x_{04}^2 (x_{03}^2 - m^2)} =
  z^{-1} d z d^3 \vec{y}
  \frac 1 {1 - \vec{y}^2}
  \frac 1 {2 x_{12} \cdot (1, \vec{y}) + z (1 - \vec{y}^2)} \\
  \frac 1 {2 x_{14} \cdot (1, \vec{y}) + z (1 - \vec{y}^2)}
  \frac 1 {x_{13}^2 + 2 z x_{13} \cdot (1, \vec{y}) + z^2 (1 - \vec{y}^2) - m^2}.
\end{multline}
Taking the residue in $z$ we find
\begin{equation}
  \operatorname{res}_{z = 0} \omega =
  \frac 1 4 \frac {d^3 \vec{y}}{(1 - \vec{y}^2) (x_{12} \cdot (1, \vec{y})) \; (x_{14} \cdot (1, \vec{y})) (x_{13}^2 - m^2)}.
\end{equation}
Next, we take a residue in the $\vec{y}^2 = 1$ we find (in a coordinate chart where $y_3 \neq 0$):

\begin{equation}
  \label{eq:box-tadpole-integrand}
  \operatorname{res}_{\vec{y}^2 = 1} \operatorname{res}_{z = 0} \omega =
  -\frac 1 {8 (x_{13}^2 - m^2)} \frac {d y_1 d y_2}{y_3 (x_{21} \cdot (1, \vec{y})) \; (x_{41} \cdot (1, \vec{y}))}.
\end{equation}
One can see that this residue cancels against a similar residue for one of the triangles in eq.~\eqref{eq:box-1mass-finite}.

Next, consider the same integral but with a squared propagator.  We have
\begin{multline}
  \omega = \frac {d^4 x_0}{(x_{01}^2)^2 x_{02}^2 x_{04}^2 (x_{03}^2 - m^2)} =\\
  \frac {d z d^3 \vec{y}}{z^3 (1 - \vec{y}^2)^2}
  \frac 1 {2 x_{12} \cdot (1, \vec{y}) + z (1 - \vec{y}^2)}
  \frac 1 {2 x_{14} \cdot (1, \vec{y}) + z (1 - \vec{y}^2)} \\
  \frac 1 {x_{13}^2 + 2 z x_{13} \cdot (1, \vec{y}) + z^2 (1 - \vec{y}^2) - m^2}.
\end{multline}
The poles at $z = 0$ and $\vec{y}^2 = 1$ are higher order so it is a bit more involved to compute the residue.  The residue at $z = 0$ is simpler since it involves a single variable.  To take the residue in $\vec{y}^2 - 1$ we parameterize
\begin{equation}
  \label{eq:two-sphere-param}
  \vec{y} = \sigma \vec{n} = \sigma \Bigl(\frac {2 t}{1 + t^2 + u^2}, \frac {2 u}{1 + t^2 + u^2}, \frac {-1 + t^2 + u^2}{1 + t^2 + u^2}\Bigr).
\end{equation}
Then, $\vec{y}^2 = \sigma^2$ so $1 - \vec{y}^2 = 1 - \sigma^2$.  We also have\footnote{The coordinates $(t, u)$ parameterize one affine patch of the unit two-sphere.  Another patch is parameterized by coordinates $(t', u')$ with $t' = \frac{t}{t^2 + u^2}$ and $u' = \frac{u}{t^2 + u^2}$ on the overlap.}
\begin{equation}
  d^3 \vec{y} = -\frac {4 \sigma^2}{(1 + t^2 + u^2)^2} d \sigma d t d u.
\end{equation}

We want to take the residue at $z = 0$ followed by the residue at $\sigma = 1$.  In these variables the differential form reads
\begin{multline}
  \omega =
  -4 \frac {d z}{z^3} \frac {d \sigma}{(1 - \sigma^2)^2} \frac {\sigma^2 d t d u}{(1 + t^2 + u^2)^2} 
  \frac 1 {2 x_{12} \cdot (1, \sigma \vec{n}) + z (1 - \sigma^2)} \\
  \frac 1 {2 x_{14} \cdot (1, \sigma \vec{n}) + z (1 - \sigma^2)}
  \frac 1 {x_{13}^2 + 2 z x_{13} \cdot (1, \sigma \vec{n}) + z^2 (1 - \sigma^2)-m^2}.
\end{multline}

Schematically, the differential form is
\begin{multline}
  \omega = -4 \frac {d z}{z^3} \frac {\sigma^2 d \sigma d t d u}{(1 - \sigma^2)^2 (1 + t^2 + u^2)^2}
  \frac 1 {A_0 + (A_1 - A_0) \sigma + z (1 - \sigma^2)} \\
  \frac 1 {B_0 + (B_1 - B_0) \sigma + z (1 - \sigma^2)} \\
  \frac 1 {C_0 + z (C_1 + \sigma (C_2 - C_1)) + z^2 (1 - \sigma^2)},
\end{multline}
with
\begin{gather}
  A_0 = 2 x_{12} \cdot (1, \vec{0}), \qquad
  A_1 = 2 x_{12} \cdot (1, \vec{n}), \\
  B_0 = 2 x_{14} \cdot (1, \vec{0}), \qquad
  B_1 = 2 x_{14} \cdot (1, \vec{n}), \\
  C_0 = x_{13}^2 - m^2, \qquad
  C_1 = 2 x_{13} \cdot (1, \vec{0}), \qquad
  C_2 = 2 x_{13} \cdot (1, \vec{n}),
\end{gather}
and we have
\begin{multline}
  \operatorname{res}_{z = 0} \operatorname{res}_{\sigma = 1} \omega =
  -4 \frac {d t d u}{(1 + t^2 + u^2)^2 C_0^2} \Bigl(
  \frac{1}{2 A_1 B_1}
  -\frac{C_2}{2 A_1^2 B_1}
  -\frac{C_2}{2 A_1 B_1^2} \\
  +\frac{A_0 C_2^2}{4 A_1^2 B_1 C_0} 
  +\frac{B_0 C_2^2}{4 A_1 B_1^2 C_0}
  -\frac{C_1 C_2}{2 A_1 B_1 C_0} 
  +\frac{C_2^2}{4 A_1 B_1 C_0}
\Bigr).
\label{eq:tadpole-residue}
\end{multline}
The only quantities depending on $\vec{n}$ are $A_1$, $B_1$ and $C_2$.

\begin{figure}
    \centering
    \includegraphics{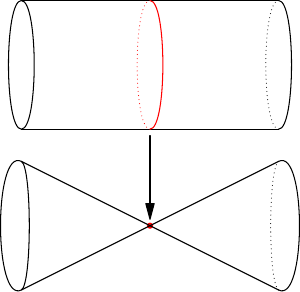}
    \caption{Blowup of the zero momentum singularity.}
    \label{fig:blowup}
\end{figure}

The pinch kinematics is now the real sphere resulting from the blowup (see fig.~\ref{fig:blowup}).  This is a two-dimensional homology class which can be identified with the unique generator of $H_2(S^2)$.  According to our conjecture, the divergence of the linear combination of integrals is equal to the integral of the residue two-form above over the real two-sphere. We remark that while these residues are computed through spherical integrals, they still preserve Lorentz symmetry; the conformal symmetry of a two-sphere is the original Lorentz symmetry of the integrals. A more detailed discussion is presented in appendix~\ref{sec:spherical-integral}

Proceeding as for the bubble permanent pinches, one can reduce the two-form in eq.~\eqref{eq:tadpole-residue} to a multiple of the two-dimensional cohomology class on the sphere, a representative of which can be taken to be $\frac{d t d u}{(1 + t^2 + u^2)^2}$.

\subsection{Squared tadpole in a bubble diagram}
The squared-tadpole divergence in the box, described in the previous section, can be canceled by adding a bubble integral $I_m(2, 0, 1, 0)$, multiplied by an appropriate coefficient.  This integral has a divergence
\begin{equation}
    \label{eq:bubble-squared-massive-div}
    \frac 1 \epsilon \frac{1}{m^2 - s_{23}}.
\end{equation}
Let us reproduce it from a double tadpole residue.

Since we have $x_{01}^2 = z^2 (1 - \vec{y}^2)$ and $x_{03}^2 = x_{13}^2 + 2 z x_{13} \cdot (1, \vec{y}) + z^2 (1 - \vec{y}^2)$, and $\vec{y} = \sigma \vec{n}$ with $\vec{n}^2 = 1$, then the integrand of $I_m(2, 0, 1, 0)$ can be written as
\begin{multline}
    \omega(2, 0, 1, 0) = \frac{d^4 x_0}{(x_{01}^2)^2 (x_{03}^2 - m^2)} =\\
    \frac{z^3 d z d \vec{y}}{z^4 (1 - \vec{y}^2)^2 (x_{13}^2 - m^2 + 2 z x_{13} \cdot (1, \vec{y}) + z^2 (1 - \vec{y}^2))} =\\
    \frac{z^3 d z \sigma^2 d \sigma d^2 \vec{n}}{z^4 (1 - \sigma^2)^2 (x_{13}^2 - m^2 + \cdots)}.
\end{multline}
Taking the residue in $z = 0$ simplifies the answer to
\begin{equation}
    \operatorname{res}_{z = 0} \omega(2, 0, 1, 0) =
    \frac{\sigma^2 d \sigma d^2 \vec{n}}{(1 - \sigma^2)^2 (x_{13}^2 - m^2)}.
\end{equation}
Finally, taking a residue at $\sigma = 1$ we obtain
\begin{equation}
    \operatorname{res}_{z = 0, \sigma = 1} \omega(2, 0, 1, 0) =
    \frac{d^2 \vec{n}}{4 (x_{13}^2 - m^2)}.
\end{equation}
If we compute the integral of this residue along the unit two-sphere, we find
\begin{equation}
    \int_{S^2} \operatorname{res}_{z = 0, \sigma = 1} \omega(2, 0, 1, 0) = \frac{\pi}{x_{13}^2 - m^2}.
\end{equation}
Up to numerical factors, this is equal to the divergent part of the bubble $I_m(2, 0, 1, 0)$ in eq.~\eqref{eq:bubble-squared-massive-div}.

We note that the cancellation described here between the integrated values of divergences is not valid at the integrand pinch level, unlike that of earlier examples. Therefore, specialized software for the computation of finite integrals, such as \texttt{trillo}~\cite{trillo} and \texttt{pySecDec}~\cite{Borowka:2017idc,Heinrich:2023til} cannot perform the integral corresponding to this linear combination, when restricted to exactly four dimensions. We leave the investigation of a locally finite form for such integrals for future work.

\subsection{Soft residue in a massive-massless exchange}
Consider the integral
\begin{equation}
  \int \frac {d^4 x_0}{x_{01}^2 (x_{02}^2 - m_2^2) (x_{03}^2 - m_3^2)},
\end{equation}
with $x_{12}^2 = m_2^2$ and $x_{13}^2 = m_3^2$.  This corresponds to the triangle integral called ``Triangle 6'' in ref.~\cite[p.~15]{Ellis:2007qk}.

This integral has a soft permanent pinch where $x_0 = x_1$.  Parameterizing as above $x_0 = x_1 + z (1, \vec{y})$ we obtain the integrand
\begin{equation}
  \omega_{\triangle} = \frac {d z d^3 \vec{y}}{z (1 - \vec{y}^2) (2 x_{12} \cdot (1, \vec{y}) + z (1 - \vec{y}^2)) (2 x_{13} \cdot (1, \vec{y}) + z (1 - \vec{y}^2))}.
\end{equation}
Taking the residue in $z$ we obtain
\begin{equation}
  \operatorname{res}_{z = 0} \omega_{\triangle} = \frac 1 4 \frac {d^3 \vec{y}}{(1 - \vec{y}^2) (x_{12} \cdot (1, \vec{y})) (x_{13} \cdot (1, \vec{y}))}.
\end{equation}
Next we use the expression of $\vec{y}$ in terms of $(\sigma, t, u)$ as in eq.~\eqref{eq:two-sphere-param}.  In these new coordinates the differential form is
\begin{equation}
  -\frac {\sigma^2 d \sigma d t d u}{(1 - \sigma^2) (1 + t^2 + u^2)^2 (x_{12} \cdot (1, \vec{y})) (x_{13} \cdot (1, \vec{y}))}.
\end{equation}
Taking the residue in $\sigma = 1$ we find
\begin{equation}
  \operatorname{res}_{\sigma = 1} \operatorname{res}_{z = 0} = \frac 1 2 \frac {d t d u}{(1 + t^2 + u^2)^2 (x_{12} \cdot (1, \vec{y})) (x_{13} \cdot (1, \vec{y}))}.
\end{equation}
Using the results of sec.~\ref{sec:spherical-integral} this can be rewritten as
\begin{equation}
  \int \frac {\delta(X^2) d^4 X}{\operatorname{Vol}(\mathbb{R}^\times) (X \cdot X_2) (X \cdot X_3)},
\end{equation}
where $X_2^2 = m_2^2$, $X_3^2 = m_3^2$.  This is actually a two-dimensional bubble integral with masses $m_2$ and $m_3$ and with $s = x_{23}^2$.  It integrates to
\begin{equation}
  \frac 1 {\sqrt{s - (m_2 - m_3)^2} \sqrt{s - (m_2 + m_3)^2}} \log \frac {\sqrt{s - (m_2 + m_3)^2} + \sqrt{s - (m_2 - m_3)^2}}{\sqrt{s - (m_2 + m_3)^2} - \sqrt{s - (m_2 - m_3)^2}}.
\end{equation}
This is the same as the $\epsilon^{-1}$ term in ref.~\cite{Ellis:2007qk}, but it is written in a slightly different form there.

Hence, the two-dimensional dual-conformal bubble integral proves to be useful for four-dimensional physics!

The residue in $z$ receives contributions from the propagator in $x_{01}$ and the propagators in $x_{02}$ and $x_{03}$, as well as the Jacobian. Similarly, the residue in $\sigma$ receives a contribution from the propagator in $x_{01}$, but also from the Jacobian. In general, we emphasize that residues are not associated with a specific set of propagators, but rather with the kinematic configuration identified by a given pinch.

\subsection{Non-squared tadpoles}
Let us end this section by confirming the claim stated in sec.~\ref{sec:simple-examples} about tadpoles not needing separate consideration, in the case of a simple box integral where no propagators are of higher powers. 
\begin{gather}
    \label{eq:box-1mass-res1}
    \operatorname{res}_{q_1} I_m(1, 1, 1, 1) = \frac{2}{A_1 B_1 C_0}, \\
    \label{eq:box-1mass-res2}
    \operatorname{res}_{q_1} I_m(1, 1, 0, 1) = \frac{2}{A_1 B_1}, \\
    \label{eq:box-1mass-res3}
    \operatorname{res}_{q_1} I_m(1, 0, 1, 1) = \operatorname{res}_{q_1} I_m(1, 1, 1, 0) = 0,
\end{gather}
where $C_0 = s_{23} - m^2$.  It follows that in the combination eq.~\eqref{eq:box-1mass-finite}, which was determined so as to cancel bubble pinches, the tadpole residue in $q_1$ also cancels.  One can check that the same happens for the tadpoles in $q_2$ and $q_4$.  Indeed, following the derivation in sec.~\ref{sec:residues} one can see that already the residue in $z$ vanishes.  The triangle residue cancels as well, so the linear combination is completely free of permanent pinches.  One can check that it is finite. 

Note that for the tadpole permanent pinch (where the momentum $q_1$ is soft, $q_1 = 0$), we have that $q_2$ and $q_4$ are also on-shell.  In this case, if one proceeds naively (by just replacing the kinematics) we actually find ourselves in the triangle kinematics as well. The computation of residues help us distinguish between these two situations; in the triangle kinematics we take four residues to obtain a zero-form, while in the tadpole case we take two residues (after a blow-up) to obtain a two-form.

%% file: double-box.tex
\section{Examples at two loops and higher}
\label{sec:two-loop-example}
\subsection{Checks on two- and three-loop topologies}
\begin{figure}[ht]
\centering
\includegraphics[width=0.58\textwidth]{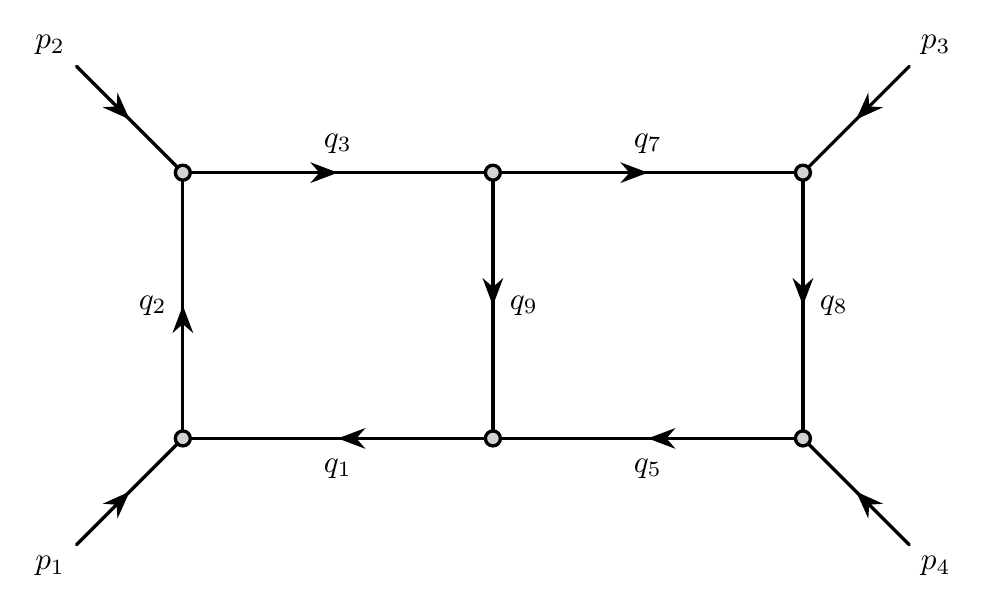}
\caption{The double-box topology with nine propagators, where the 4th and 6th are the irreducible scalar products (ISP).}
\label{fig:double-box}
\end{figure}

We illustrate the idea of using permanent pinches to determine finite integrals in the double-box topologies studied in refs.~\cite{Wasser:2018qvj,Gambuti:2023eqh}, as shown in fig.~\ref{fig:double-box}. We found that imposing the cancellation of one-loop and two-loop pinches on a generic ansatz of rank two leads to finite integrals, where the rank denotes the degree in the loop momenta $\ell_i$ of any monomial in the numerator. Making use of the notation in ref.~\cite{Henn:2020lye}, the topology is defined by the following nine propagators
\begin{align}
\label{eq:db_props}
D_1 &= k_1^2, &
D_2 &= (k_1+p_1)^2, &
D_3 &= (k_1+p_1+p_2)^2, \notag\\
D_4 &= (k_1+p_1+p_2+p_3)^2, &
D_5 &= k_2^2, &
D_6 &= (k_2+p_1)^2, \notag\\
D_7 &= (k_2+p_1+p_2)^2, &
D_8 &= (k_2+p_1+p_2+p_3)^2, &
D_9 &= (k_1-k_2)^2.
\end{align}

\begin{figure}[ht]
\centering
\begin{subfigure}[b]{0.46\textwidth}
\centering
\vbox to 4.95cm{\vfil
\includegraphics[width=\textwidth]{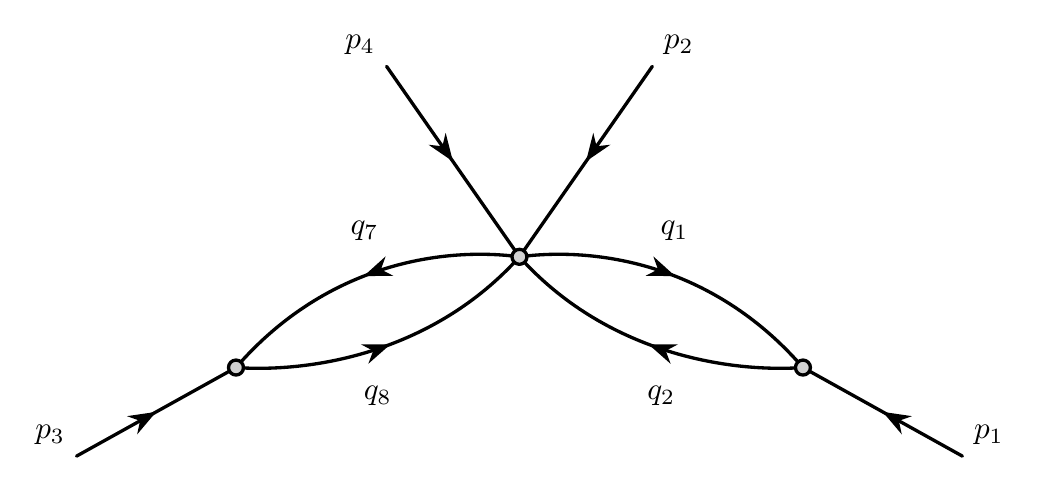}
\vfil}
\caption{}
\end{subfigure}
\hspace{0.02\textwidth}
\begin{subfigure}[b]{0.46\textwidth}
\centering
\vbox to 4.95cm{\vfil
\includegraphics[width=0.88\textwidth]{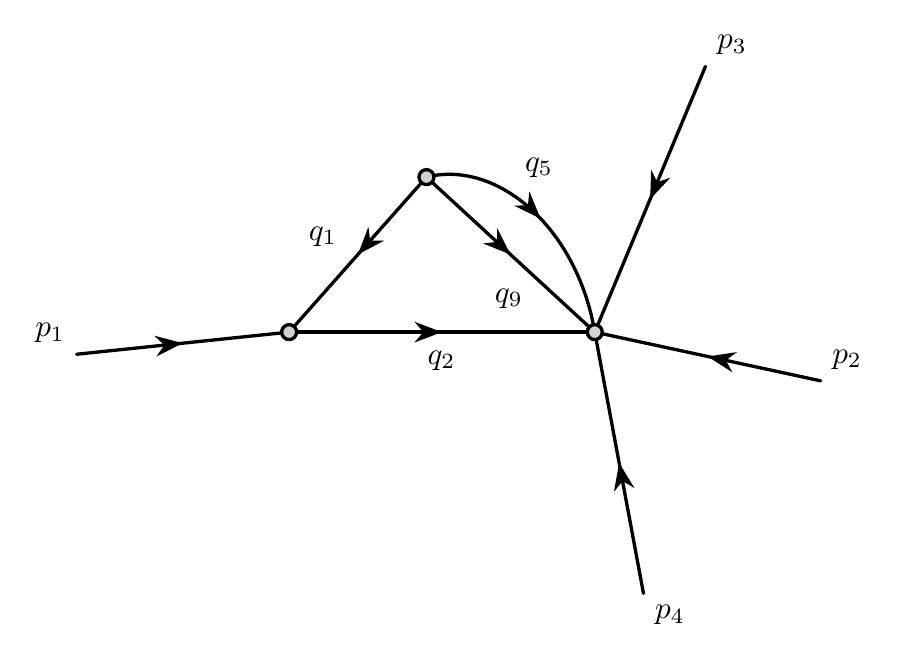}
\vfil}
\caption{}
\end{subfigure}
\caption{Two-loop pinches of the double-box topology. The diagrams are drawn with the help of the program Azurite~\cite{Georgoudis:2016wff}.
}
\label{fig:double-box-subtopologies}
\end{figure}

We first show the finite integrals in the literature~\cite{Wasser:2018qvj} are vanishing at the pinch kinematics. One of the two finite double box integrals in~\cite{Wasser:2018qvj} is
\begin{align} \label{eq:b2}
 &F_1=
 t\,\jint{0,1,1,0,1,0,0,1,1}
- s\,\jint{1,0,1,0,1,0,1,0,1}
+ t\,\jint{1,1,0,0,0,0,1,1,1} \\
&+ s\,\jint{1,1,1,-1,1,-1,1,1,1}
- st\,\jint{1,1,1,0,1,0,1,1,0}\,,
\end{align}
where $s=(p_1+p_2)^2$ and $t=(p_2+p_3)^2$.
At the two-loop pinch in fig.~\ref{fig:double-box-subtopologies} (a), the propagator momenta $q_i$ ($D_i=q_i^2$) can be parameterized as
\begin{align}
q_1 &\to -(1-z)p_1, &
q_2 &\to z p_1, \notag\\
q_7 &\to -(1-w)p_3, &
q_8 &\to w p_3\,,
\end{align}
where the four propagators become collinear with either $p_1$ or $p_3$, and therefore $D_i\to 0$ for $i=1,2,7,8$.
According to momentum conservation, other propagators read 
\begin{align}
D_3 &\to sz, &
D_4 &\to t + sz - (s+t)z, \notag\\
D_5 &\to (s+t)w - tw, &
D_6 &\to t - tw, \notag\\
D_9 &\to t - tw + sz - (s+t)z + (s+t)wz.
\end{align}
The integral 
$F_1$ at this kinematics becomes
\begin{align}
F_1&\to\frac{t}{D_9}
+ s\,\frac{D_4D_6}{D_3D_5D_9}
- st\,\frac{1}{D_3D_5}\nonumber\\
&=\frac{t}{t(1-w)(1-z)+swz}
+\frac{t^2(1-w)(1-z)}
{s\,wz\left[t(1-w)(1-z)+swz\right]}
-\frac{t}{s\,wz}=0\,,
\end{align}
where only the last three terms in eq.~\eqref{eq:b2} have non-zero contributions, and they sum up to zero as expected. We also checked $F_1$ is vanishing on the pinch kinematics depicted in fig.~\ref{fig:double-box-subtopologies} (b).

Let us show how a leading IR divergence can be computed in the example of the first term in the combination $F_1$, which in the pinch kinematics of fig.~\ref{fig:double-box-subtopologies}, subfigure (a), becomes
\begin{equation}
    \jint{0,1,1,0,1,0,0,1,1}\rightarrow\frac{t}{t (1 - z) (1 - w) + s z w}.\label{eq:slashed-box-integral}
\end{equation}
\begin{figure}
    \centering
    \includegraphics[width=0.25\linewidth]{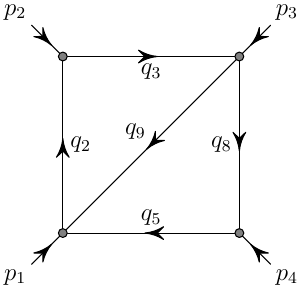}
    \caption{Two-loop slashed-box integral appearing in eq.~\eqref{eq:slashed-box-integral}}
    \label{fig:slashed-box}
\end{figure}
This is a two-loop slashed box integral (see fig.~\ref{fig:slashed-box}).  One can check that the permanent pinch in fig.~\ref{fig:double-box-subtopologies} (a) is the only leading permanent pinch of the slashed box. We can therefore predict that the leading IR divergence of this integral is
\begin{equation}
    \int_0^1 d z \int_0^1 d w \frac{t}{t (1 - z) (1 - w) + s z w} = \frac{t}{2 (s + t)} \Bigl(\log^2 \frac{s}{t} + \pi^2\Bigr).
\end{equation}
This can be confirmed to be correct by a calculation of the $\epsilon^{-2}$ coefficient of the expansion around $D = 4$ of the dimensional-regulated integral~\cite{Henn:2013pwa}.

\begin{figure}[ht]
\centering
\includegraphics[width=0.48\textwidth]{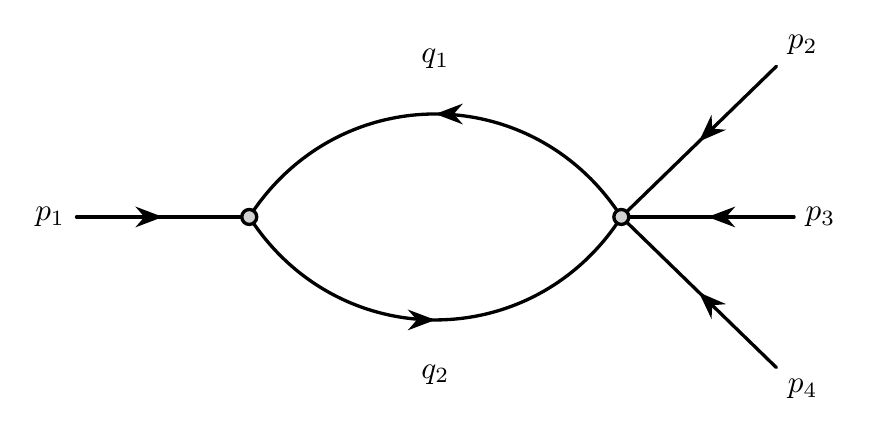}
\caption{One-loop pinch of the double-box topology.}
\label{fig:one_loop_pinch}
\end{figure}

We next look at the one-loop pinch shown in fig.~\ref{fig:one_loop_pinch}, where the kinematics is 
\begin{align}
q_1 &\to -(1-z)p_1, &
q_2 &\to z p_1\,,
\end{align}
which implies $D_1\to 0, D_2\to 0$. According to momentum conservation, the remaining seven propagators are given by
\begin{align}
D_3 &\to sz, &
D_4 &\to t(1-z), \notag\\
D_5 &\to w, &
D_6 &\to w+2w_1, \notag\\
D_7 &\to s+w+2(w_1+w_2), &
D_8 &\to w+2(w_1+w_2+w_3), \notag\\
D_9 &\to w+2w_1(1-z)\,,
\end{align}
where $w=q_5^2$ and $w_i=q_5\cdot p_i$ for $i=1,2,3$. Only the last three terms in eq.~\eqref{eq:b2} have non-zero contributions, 
\begin{align}
F_1
&\to \frac{t}{D_7D_8D_9}
+s\,\frac{D_4D_6}{D_3D_5D_7D_8D_9}
-st\,\frac{1}{D_3D_5D_7D_8}\nonumber\\
&=
-\frac{t}
{\bigl(s+w+2w_1+2w_2\bigr)
 \bigl(w+2w_1+2w_2+2w_3\bigr)
 \bigl(-w+2w_1(-1+z)\bigr)}
\nonumber\\
&\quad
+\frac{t\,(w+2w_1)(-1+z)}
{w\bigl(s+w+2w_1+2w_2\bigr)
 \bigl(w+2w_1+2w_2+2w_3\bigr)
 \bigl(-w+2w_1(-1+z)\bigr)z}
\nonumber\\
&\quad
-\frac{t}
{w\bigl(s+w+2w_1+2w_2\bigr)
 \bigl(w+2w_1+2w_2+2w_3\bigr)z}=0\,,
\end{align}
and they again sum up to zero as expected.

\begin{figure}[ht]
\centering
\includegraphics[width=0.62\textwidth]{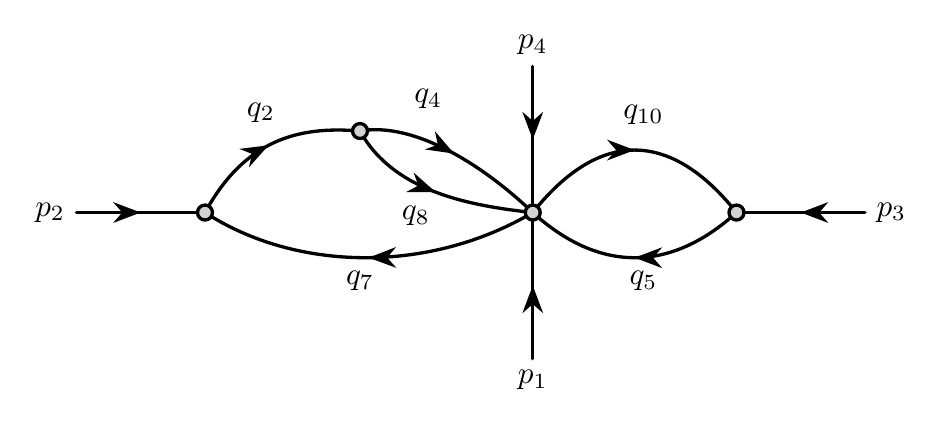}
\caption{Pinch topology of the triple-box integral.}
\label{fig:triple-box-pinch}
\end{figure}

Starting with a generic ansatz for numerators in double-box topology containing terms up to products of \emph{two} propagators, which contains $26$ integrals, by imposing both the one-loop and two-loop pinches, the coefficients multiplying $22$ of them are fixed (up to a global scale), resulting in four independent finite integrals. Two of those contain the top sector topology: these are
\begin{align}
F_{\text{db},1}
={}& \frac{t}{s}\,\jint{0,1,1,0,1,0,0,1,1}
+ \frac{t}{s}\,\jint{1,1,0,0,0,0,1,1,1}
+ \jint{1,1,1,-1,1,-1,1,1,1}
- t\,\jint{1,1,1,0,1,0,1,1,0}\,,
\\[0.6em]
F_{\text{db},2}
={}& \frac{t}{s}\jint{0,1,1,0,0,0,1,1,1}
   + \jint{0,1,1,0,1,-1,1,1,1}
   - \Bigl(1-\frac{t}{s}\Bigr) \jint{0,1,1,0,1,0,0,1,1}
 \notag\\
&- t\,\jint{0,1,1,0,1,0,1,1,1}
   - \Bigl(1-\frac{t}{s}\Bigr)\jint{1,1,0,0,0,0,1,1,1}
   + \jint{1,1,0,0,1,-1,1,1,1} \notag\\
&+ \frac{t}{s}\jint{1,1,0,0,1,0,0,1,1}
   - t\,\jint{1,1,0,0,1,0,1,1,1}
   + \jint{1,1,1,-1,0,0,1,1,1}
   + \jint{1,1,1,-1,1,0,0,1,1} \notag\\
&- s\,\jint{1,1,1,-1,1,0,1,1,1}
   - t\,\jint{1,1,1,0,0,0,1,1,1}
   - s\,\jint{1,1,1,0,1,-1,1,1,1}
   - t\,\jint{1,1,1,0,1,0,0,1,1} \notag\\
&+ s\,\jint{1,1,1,0,1,0,1,1,0}
   + st\,\jint{1,1,1,0,1,0,1,1,1}\,.
\end{align}
which we verified to map to linear combinations of the two integrals of this type described in~\cite{Wasser:2018qvj}.
We can extend the analysis to three loops. Let us consider, for example, a finite integral described in~\cite{Henn:2020lye} for the three-loop triple-box, given by the expression
\begin{figure}[ht]
\centering
\includegraphics[width=0.62\textwidth]{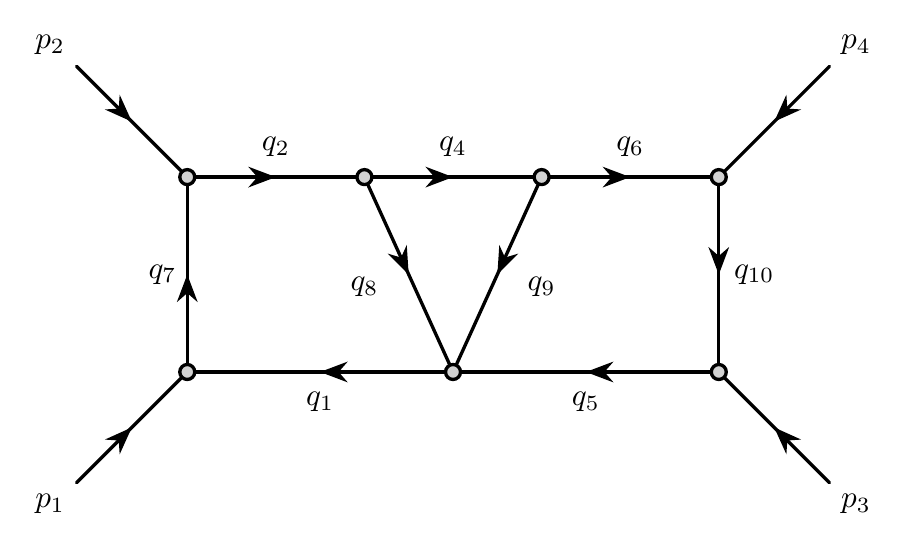}
\caption{Triple-box topology with one propagator contracted.}
\label{fig:triple-box-contracted}
\end{figure}
\begin{equation}
\begin{aligned}
F_{\text{tb}}
={}& -s_{12}s_{13}\,\jint{1,1,0,1,1,1,1,1,1,0,0,0,0,-1}
-s_{12}\,\jint{1,1,0,1,1,1,0,1,1,0,0,0,0,0}
\\
&+s_{12}\,\jint{1,1,0,1,1,1,1,1,1,-1,0,0,-1,0}
+s_{13}\,\jint{0,1,0,1,1,0,1,1,1,1,0,0,0,0}
\\
&+s_{13}\,\jint{1,0,0,1,0,1,1,1,1,1,0,0,0,0}\,.
\label{eq:finiteA8}
\end{aligned}
\end{equation}
As an example, we study the following pinch, which is a product of two-loop and one-loop pinches as shown in fig.~\ref{fig:triple-box-pinch},
with the parameterization in the collinear limit
\begin{align}
q_2 &\to z\,p_2, &
q_7 &\to -(1-z)\,p_2, \nonumber\\
q_4 &\to wz\,p_2, &
q_8 &\to (1-w)z\,p_2, \nonumber\\
q_5 &\to y\,p_3, &
q_{10} &\to -(1-y)\,p_3\,
\end{align}
and the remaining propagators become
\begin{align}
D_1 &\to s_{12}(1-z), &
D_6 &\to s_{12}(1-y), &
D_9 &\to s_{13}wyz+s_{12}(y-1)(wz-1), \notag\\
D_{11} &\to s_{13}z, &
D_{14} &\to s_{13}y, &
D_{15} &\to s_{13}yz-s_{12}(y+z-yz-1)\,.
\end{align}

\begin{figure}[ht]
\centering
\begin{subfigure}[b]{0.37\textwidth}
\centering
\includegraphics[width=\textwidth]{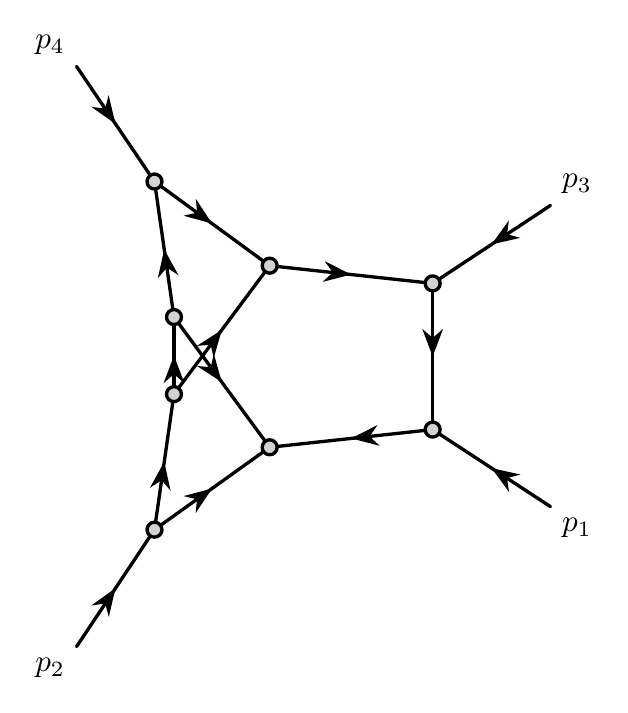}
\caption{}
\end{subfigure}
\hspace{0.01\textwidth}
\begin{subfigure}[b]{0.45\textwidth}
\centering
\includegraphics[width=\textwidth]{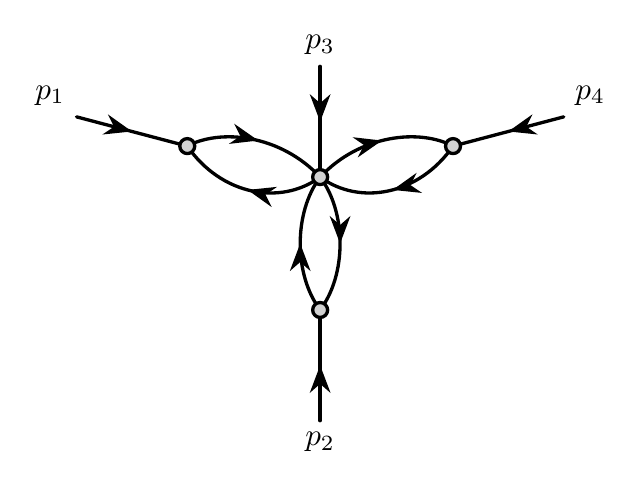}
\caption{}
\end{subfigure}
\caption{Three-loop non-planar topology and one of its pinches, where the finite integrals found in~\cite{Henn:2020lye} are vanishing.}
\label{fig:three-loop-nonplanar-pinches}
\end{figure}

In the pinch kinematics, only the 1st, 3rd, and 4th terms in eq.~\eqref{eq:finiteA8} have non-vanishing contribution on the pinch, and we have 
\begin{align}
F_{\text{tb}}
\to{}&
-s_{12}s_{13}\,\frac{D_{15}}{D_1D_6D_9}
- s_{12}\,\frac{D_{11}D_{14}}{D_1D_6D_9}
- s_{13}\,\frac{1}{D_9}
\notag\\
={}&
-\frac{s_{13}\bigl(s_{12}(y-1)(z-1)+s_{13}yz\bigr)}
{s_{12}(y-1)(z-1)\bigl(s_{13}wyz+s_{12}(y-1)(wz-1)\bigr)}
\notag\\
&+\frac{s_{13}^{2}yz}
{s_{12}(y-1)(z-1)\bigl(s_{13}wyz+s_{12}(y-1)(wz-1)\bigr)}
+\frac{s_{13}}
{s_{13}wyz+s_{12}(y-1)(wz-1)}
\notag\\
={}&0\,.
\end{align}
Similar cancellations occur for all other pinches of this topology.
We have also studied the pinches of three-loop non-planar families as shown in fig.~\ref{fig:three-loop-nonplanar-pinches}, and we found that the finite integrals described in~\cite{Henn:2020lye} for this topology vanish at those pinches. 

\begin{figure}
\centering
\includegraphics[width=0.4\textwidth]{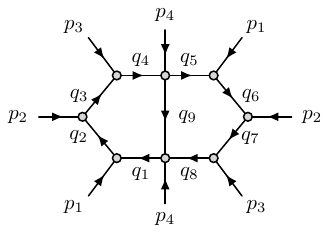}
\caption{The relevant topology for the computation of the Wilson loop observable in eq.~\eqref{eq:observable},  with nine propagators defined in eq.~\eqref{eq:db_props}.}
\label{fig:dbox_observable}
\end{figure}

More interestingly, we can also consider IR-finite observables expressible as finite integrals, e.g.\ the Wilson loop with Lagrangian insertions in $\mathcal{N}=4$ super Yang-Mills~\cite{Alday:2012hy,Henn:2019swt} and its recent negative geometry expansion~\cite{Arkani-Hamed:2021iya,Dixon:2026ipt}. The two-loop corrections can be described by the massless double-box topology (see fig.~\ref{fig:double-box}) but with slightly generalized form as shown in fig.~\ref{fig:dbox_observable}. These observables are related to the `logarithm' of the scattering amplitudes, which has non-planar structure in the dual momentum space. A double-covering of four-point kinematics is a convenient way of expressing the associated integral; nevertheless, the permanent pinches we have seen can be studied in the same way, and we found the IR-finite observable, (4.6) in~\cite{Arkani-Hamed:2021iya} as the following

{\small
\begin{align}
W
={}&
t^2\jint{0,1,1,1,0,1,0,1,1}
+st\jint{0,1,1,1,1,0,0,1,1}
+t^2\jint{0,1,1,1,1,1,-1,1,1}
+st\jint{0,1,1,1,1,1,0,0,1}
\notag\\
&+st^2\jint{0,1,1,1,1,1,0,1,1}
+st\jint{1,0,1,1,0,1,1,0,1}
+s^2\jint{1,0,1,1,1,0,1,0,1}
+st\jint{1,0,1,1,1,1,0,0,1}
\notag\\
&+s^2\jint{1,0,1,1,1,1,1,-1,1}
+s^2t\jint{1,0,1,1,1,1,1,0,1}
+t^2\jint{1,1,0,1,-1,1,1,1,1}
+st\jint{1,1,0,1,0,0,1,1,1}
\notag\\
&+t^2\jint{1,1,0,1,0,1,0,1,1}
+st\jint{1,1,0,1,0,1,1,0,1}
+st^2\jint{1,1,0,1,0,1,1,1,1}
+st\jint{1,1,1,0,0,0,1,1,1}
\notag\\
&+s^2\jint{1,1,1,0,1,-1,1,1,1}
+st\jint{1,1,1,0,1,0,0,1,1}
+s^2\jint{1,1,1,0,1,0,1,0,1}
+s^2t\jint{1,1,1,0,1,0,1,1,1}
\notag\\
&+st^2\jint{1,1,1,1,0,1,1,1,0}
+s^2t\jint{1,1,1,1,1,0,1,1,0}
+st^2\jint{1,1,1,1,1,1,0,1,0}
+s^2t\jint{1,1,1,1,1,1,1,0,0}
\notag\\
&+s^2t^2\jint{1,1,1,1,1,1,1,1,0}\,,
\label{eq:observable}
\end{align}
}
is vanishing on the pinch kinematics at one (see fig.~\ref{fig:one_loop_pinch}) and two loops (see fig.~\ref{fig:double-box-subtopologies}).  
\subsection{Finite integrals for the non-planar double pentagon}
\begin{figure}
    \centering
    \includegraphics[width=0.36\linewidth]{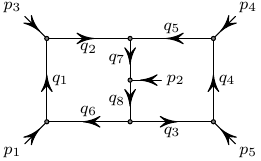}
    \caption{Graph of the non-planar `double pentagon' topology at two loops.}
    \label{fig:double-pentagon}
\end{figure}

A more elaborate example of the application of our method is given by a non-planar `double pentagon' diagram at two loops (see fig.~\ref{fig:double-pentagon}). It is an interesting question to determine all the finite integrals that are allowed in this topology.
The analysis of contracted graphs present for this topology reveals that it has a minimal set of five one-loop and $14$ two-loop bubble-type pinches whose cancellation we need to impose in order to obtain a finite integral. Starting from a generic combination of UV-finite numerators expressed in terms of scalar products containing loop momenta, excluding evanescent contributions, i.e. those whose integrand vanishes in four dimensions, we can identify a total of $3052$ linearly independent combinations that correspond to convergent integrals. We include their explicit expressions in the ancillary files. 
In appendix~\ref{sec:nonplpent} we briefly describe the procedure and conventions we use for the results. 
The finiteness of the integrals have been explicitly checked using FIRE~\cite{Smirnov:2019qkx,Smirnov:2025prc} and {\tt AMFlow} \cite{Liu:2017jxz,Liu:2022chg,Huang:2026rjb} in a generic five-point kinematics to verify the $1/\epsilon^p$ pieces for $p=1,...,4$ are canceled.  
The result presented is, to
the best of our knowledge, the first for a basis for finite integrals in a non-planar five-point topology, and partially extends recent work in ref.~\cite{Figueiredo:2026ksz}, where bases of finite integrals of this type were obtained for the planar pentabox topology. We note that with respect to refs.~\cite{Gambuti:2023eqh,Figueiredo:2026ksz}, the pinches are here determined diagrammatically rather than through solutions of the Landau equations, and the finiteness conditions are
imposed as residues on the pinch loci, evaluated over finite fields; the resulting constraints agree for the logarithmic singularities relevant to this topology, while the residue formulation extends to the higher-order poles and massive configurations of sec.~\ref{sec:residues}. A description and full implementation of the algorithm used to find the most general set of finite integrals in this topology, as well as more complicated ones, e.g. three-loop five-point integrals \cite{Chicherin:2025mvc}, is deferred to future work.

%% file: thanks.tex
\section*{Acknowledgments}

We thank
Giulio Gambuti,
Johannes Henn,
Jungwon Lim,
Elia Mazzucchelli,
Prashanth Raman,
Lorenzo Tancredi,
Jaroslav Trnka and Chen-Yu Wang
for stimulating discussions.
This project was supported by the European Research Council (ERC) under the European Union’s Horizon Europe research and innovation programme (grant agreement No.
101097219, UNIVERSE+), and by the Deutsche Forschungsgemeinschaft (DFG, German Research Foundation) through the research unit FOR 5582 – Projektnummer 508889767. Views and opinions expressed are however those of the authors only and do not necessarily reflect those of the European Union or the European Research Council Executive Agency. Neither the European Union nor the granting authority can be held responsible for them.

%% file: higher-residues.tex
\section{Notions from the theory of residues in multiple variables}
\label{sec:higher-residues}

Below, we give a quick introduction to Leray's theory of residues.  Consider a complex manifold $X$ and a hypersurface $S$ in $X$ with global equation $s(x) = 0$.  In applications we sometimes want to allow $S$ to have singularities, as in the case of the light-cone.

We will make use of a number of results (see refs.~\cite{BSMF_1959__87__81_0, phamsingularities})
\begin{lemma}
  \label{lemma:division}
  If $\omega$ is a regular (no singularities) differential form on $X$ and $d s \wedge \omega = 0$, then there exists a regular form $\psi$ on $X$ such that $\omega = d s \wedge \psi$.  We will use the notation $\psi\vert_S = \left. \frac {\omega}{d s}\right\rvert_S$ for its restriction to $S$.  This restriction depends only on $\omega$ and the equation $s(x) = 0$ and is holomorphic if $\omega$ is holomorphic.
\end{lemma}

Given a closed, regular form $\omega$ such that $d s \wedge \omega = 0$, using lemma~\ref{lemma:division} we can find $\omega_1$ such that $\omega = d s \wedge \omega_1$.  Since $\omega$ is closed we have that $d s \wedge d \omega_1 = 0$ so, applying the lemma again we can find $\omega_2$ such that $d \omega_1 = d s \omega_2$ and so on.  Since we have
\begin{equation}
  \omega_1 = \frac {\omega}{d s}, \qquad
  \omega_2 = \frac {d \omega_1}{d s}, \ldots,
\end{equation}
we will use the notation $\omega_k\vert_S = \left.\frac {d^{k - 1} \omega}{d s^k}\right\rvert_S$.

Next, consider a closed differential form $\phi$ which has a pole of order $k$ along $S$.  This means that $\omega = s^k \phi$ is regular on $X$ but $s^p \phi$ for $p < k$ is not.  We will define a notion of residue for these forms.  If $\phi$ has a first order pole, then it can be written as
\begin{equation}
  \phi = \frac {d s} s \wedge \psi + \chi,
\end{equation}
for some regular differential forms $\psi$ and $\chi$.  We define the \emph{residue form} of $\phi$ to be
\begin{equation}
  \operatorname{res} \phi = \psi\vert_S = \left.\frac {s \phi}{d s}\right\rvert_S.
\end{equation}
If $\phi$ is meromorphic then its residue is holomorphic.

We have the following important result:
\begin{lemma}
  The cohomology class $[\operatorname{res} \phi] \in H^*(S)$ depends only on the cohomology class $[\phi] \in H^{* + 1}(X \setminus S)$.  This cohomology class is often denoted by $\operatorname{Res} \phi = [\operatorname{res} \phi]$ (note the capitalization).
\end{lemma}

Let us now extend these results to the case of higher order poles.  We now have $\omega = s^k \phi$ is regular and closed and we will also impose that $d s \wedge \phi = 0$ which implies that $d \omega = 0$.  Then we define the residue to be
\begin{equation}
  \label{eq:higher-pole-residue-def}
  \operatorname{res} \phi = \frac 1 {(k - 1)!} \left.\frac {d^{k - 1} (s^k \phi)}{d s^k}\right\rvert_S.
\end{equation}

This can be extended to iterated residues as follows.  Suppose $\phi$ is a differential form with poles of order $k_1, \dotsc, k_m$ along hypersurfaces $S_1, \dotsc, S_m$ (with global equations $s_1, \dotsc, s_m$).  Then $\omega = s_1^{k_1} \cdots s_m^{k_m} \phi$ is regular.  Then,
\begin{equation}
  \operatorname{res}^m \phi = \frac 1 {(k_1 - 1)! \cdots (k_m - 1)!} \left.\frac {d^{k_1 + \cdots + k_m - m} (s_1^{k_1} \cdots s_m^{k_m} \phi)}{d s_1^{k_1} \wedge \cdots \wedge d s_m^{k_m}}\right\rvert_{S_1 \cap \cdots \cap S_m}.
\end{equation}

Let us work out an example where $\omega = f(x, y) d x d y$, $s = \frac 1 2 (x^2 - y^2)$ with $d s = x d x - y d y$.

We have $\omega = d s \omega_1$ for $\omega_1 = \frac {d y} x f(x, y)$ or
$\omega_1 = \frac {d x} y f(x, y)$, but let's pick the first option.  The restriction to $S$ of these two one-forms is the same since on $S$ we have $x d x = y d y$.

Then,
\begin{equation}
  d \omega_1 = \frac {(x d x - y d y)}{x} \Bigl(\frac {\partial_x f}{x} - \frac f {x^2}\Bigr) d y
\end{equation}
so
\begin{equation}
  \omega_2 = \Bigl(\frac {\partial_x f}{x^2} - \frac f {x^3}\Bigr) d y.
\end{equation}

Next,
\begin{equation}
  \label{eq:omega2}
  d \omega_2 = \frac {x d x - y d y}x \Bigl(\frac {\partial_x^2 f}{x^2} - 3 \frac {\partial_x f}{x^3} + 3 \frac f {x^4}\Bigr) d y,
\end{equation}
so
\begin{equation}
  \label{eq:omega3}
  \omega_3 = \Bigl(\frac {\partial_x^2 f}{x^3} - 3 \frac {\partial_x f}{x^4} + 3 \frac f {x^5}\Bigr) d y
\end{equation}
and so on.

Suppose we have a differential form with higher order pole
\begin{equation}
  \phi = \frac {d x d y f(x, y)}{s^k}
\end{equation}
and $\omega = s^k \phi = f(x, y) d x d y$.  On one hand, we can take the residue easily by changing variables to $\tilde{z} = \frac 1 2 (x + y)$, $z = \frac 1 2 (x - y)$, $x = z + \tilde{z}$, $y = -z + \tilde{z}$ and $s = 2 z \tilde{z}$, $d x \wedge d y = 2 d z \wedge d \tilde{z}$ so
\begin{equation}
  \phi = \frac 1 {2^{k - 1}} \frac {d z d \tilde{z} F(z, \tilde{z})}{z^k \tilde{z}^k},
\end{equation}
where $F(z, \tilde{z}) = f(z + \tilde{z}, -z + \tilde{z})$.

Then, we have that
\begin{equation}
  \operatorname{res}_{z = \tilde{z} = 0} \frac {d z d \tilde{z} F(z, \tilde{z})}{z^k \tilde{z}^k} = \frac 1 {((k - 1)!)^2} \partial_z^{k - 1} \partial_{\tilde{z}}^{k - 1} F(0, 0).
\end{equation}
Using $\partial_z = \partial_x - \partial_y$ and $\partial_{\tilde{z}} = \partial_x + \partial_y$ we find
\begin{equation}
  \operatorname{res}_{x = y = 0} \frac {d x d y f(x, y)}{s^k} =
  \frac 1 {2^{k - 1}} \frac 1 {((k - 1)!)^2} \sum_{l = 0}^{k - 1} (-1)^l \binom{k - 1}{l} \partial_x^{2 l} \partial_y^{2 k - 2 l - 2} f(0, 0).
\end{equation}
The first few expressions read
\begin{gather}
  \operatorname{res}_{x = y = 0} \frac {d x d y f(x, y)}{s} = f(0, 0), \\
  \operatorname{res}_{x = y = 0} \frac {d x d y f(x, y)}{s^2} = \frac 1 2 (\partial_x^2 - \partial_y^2) f(0, 0), \\
  \operatorname{res}_{x = y = 0} \frac {d x d y f(x, y)}{s^3} = \frac{1}{16} (\partial_x^4 - 2 \partial_x^2 \partial_y^2 + \partial_y^4) f(0, 0).
\end{gather}

On another hand, we can take the residue in $s = 0$ as described above and then take the residue in one of the variables $x$ or $y$ which are now linked by the $s = 0$ equation.  Taking the residue in $s = 0$ we have
\begin{equation}
  \operatorname{res} \frac {d x d y f(x, y)}{s^k} =
  \frac 1 {(k - 1)!} \frac {d^{k - 1} (f(x, y) d x d y)}{d s^k}.
\end{equation}
In particular, we have
\begin{gather}
  \operatorname{res} \frac {d x d y f(x, y)}{s} = \frac {f(x, y) d x d y}{d s} = \omega_1, \\
  \operatorname{res} \frac {d x d y f(x, y)}{s^2} = \frac {d (f(x, y) d x d y)}{d s^2} = \omega_2, \\
  \operatorname{res} \frac {d x d y f(x, y)}{s^3} = \frac 12\frac {d^2 (f(x, y) d x d y)}{d s^3} = \frac 1 2 \omega_3.
\end{gather}

The differential forms $\omega_k$ above naturally live on the hypersurface $S$.  On this hypersurface we can pick coordinates $y = \pm x$ (so we have two branches of solutions as in some similar spinor computations).

Taking $y = x$ for $\omega_2$ in eq.~\eqref{eq:omega2}, we find
\begin{equation}
  (\partial_x + \partial_y) \partial_x f -
  (\frac 1 2 \partial_x^2 + \partial_x \partial_y + \frac 1 2 \partial_y^2) f =
  \frac 1 2 (\partial_x^2 - \partial_y^2) f(0, 0).
\end{equation}

Taking $y = x$ for $\omega_3$ in eq.~\eqref{eq:omega3}, we find
\begin{multline}
  \Bigl(\frac 1 2 \partial_x^2 + \partial_x \partial_y + \frac 1 2 \partial_y^2\Bigr) \partial_x^2 f \\
  -3 \Bigl(\frac 1 {3!} \partial_x^3 + \frac 1 2 \partial_x^2 \partial_y + \frac 1 2 \partial_x \partial_y^2 + \frac 1 {3!} \partial_y^3\Bigr) \partial_x f + \\
  3 \Bigl(\frac 1 {4!} \partial_x^4 + \frac 1 {3!} \partial_x^3 \partial_y + \frac 1 {2! 2!} \partial_x^2 \partial_y^2 + \frac 1 {3!} \partial_x \partial_y^3 + \frac 1 {4!} \partial_y^4\Bigr) f = \\
  \frac 1 {8} (\partial_x^4 - 2 \partial_x^2 \partial_y^2 + \partial_y^4) f(0, 0).
\end{multline}
In fact, since the derivatives of odd order in $y$ cancel, we obtain the same answer if we pick the other solution $y = -x$ in $f$.

%% file: spherical-integral.tex
\section{A spherical integral}
\label{sec:spherical-integral}

When taking tadpole residues in sec.~\ref{sec:squared-tadpoles}, we obtain integrals of type
\begin{equation}
  J(n_a, n_b, n_c) = \int_{S^2} \frac {d^2 \vec{n}}{(a_0 + \vec{n} \cdot \vec{a})^{n_a} (b_0 + \vec{n} \cdot \vec{b})^{n_b} (c_0 + \vec{n} \cdot \vec{c})^{n_c}},
\end{equation}
where the integral is over the unit sphere. We introduce stereographic coordinates
\begin{equation}
    \vec{n} = \Bigl(\frac{2 t}{1 + t^2 + u^2}, \frac{2 u}{1 + t^2 + u^2}, \frac{-1 + t^2 + u^2}{1 + t^2 + u^2}\Bigr),
\end{equation}
where the measure of integration is
\begin{equation}
    d^2 \vec{n} = \frac{4 d t d u}{(1 + t^2 + u^2)^2}.
\end{equation}

Using the expression of $\vec{n}$ in terms of stereographic coordinates we have
\begin{equation}
    a_0 + \vec{a} \cdot \vec{n} =
    \frac{a_0 + a_3}{1 + t^2 + u^2} \Bigl((t + \frac {a_1}{a_0 + a_3})^2 + (u + \frac {a_2}{a_0 + a_3})^2 + \frac {a^2}{(a_0 + a_3)^2}\Bigr),
\end{equation}
and similarly for $b_0 + \vec{b} \cdot \vec{n}$, etc.  Here we have introduced the notation $a^2 = a_0^2 - a_1^2 - a_2^2 - a_3^2$, etc.  In particular, in the context of calculations performed in sec.~\ref{sec:squared-tadpoles}, we have $a_0 = 2 x_{12}^0$ and $\vec{a} = 2 \vec{x}_{12}$, so that $a^2 = 0$.  Similarly, $b_0 = 2 x_{14}^0$ and $\vec{b} = 2 \vec{x}_{14}$ so $b^2 = 0$.

The integral now reads
\begin{multline}
    \frac{1}{(a_0 + a_3)^{n_a} (b_0 + b_3)^{n_b} (c_0 + c_3)^{n_c}} \int_{\mathbb{R}^2} 4 d t d u (1 + t^2 + u^2)^{-2 + n_a + n_b + n_c} \\ \Bigl((t + \frac{a_1}{a_0 + a_3})^2 + (u + \frac{a_2}{a_0 + a_3})^2\Bigr)^{-n_a} \\
    \Bigl((t + \frac{b_1}{b_0 + b_3})^2 + (u + \frac{b_2}{b_0 + b_3})^2\Bigr)^{-n_b} \\
    \Bigl((t + \frac{c_1}{c_0 + c_3})^2 + (u + \frac{c_2}{c_0 + c_3})^2 + \frac{c^2}{(c_0 + c_3)^2}\Bigr)^{-n_c}.
\end{multline}

It is convenient to make a four-vector
\begin{equation}
    X = (X^{+}, X^1, X^2, X^{-}) = (1, -t, -u, t^2 + u^2)
\end{equation}
from the coordinates $t$, $u$.  On this vector space we introduce a scalar product
\begin{equation}
    X \cdot Y = -\frac 1 2 (X^{+} Y^{-} + X^{-} Y^{+}) + X^1 Y^1 + X^2 Y^2.
\end{equation}
We have $X \cdot X = 0$.  If we introduce
\begin{gather}
    X_1 = (1, 0, 0, 1), \\
    X_2 = \Bigl(a_0 + a_3, a_1, a_2, \frac{a_1^2 + a_2^2}{a_0 + a_3}\Bigr), \\
    X_3 = \Bigl(b_0 + b_3, b_1, b_2, \frac{b_1^2 + b_2^2}{b_0 + b_3}\Bigr), \\
    X_4 = \Bigl(c_0 + c_3, c_1, c_2, c_0 - c_3\Bigr).
\end{gather}
We can write $X_2$ and $X_3$ in a form similar to $X_4$ if we use the fact that $a^2 = 0$ and $b^2 = 0$.

Then, we have
\begin{gather}
    -2\; X \cdot X_1 = 1 + t^2 + u^2, \\
    -2\; X \cdot X_2 = (a_0 + a_3) \Bigl((t + \frac{a_1}{a_0 + a_3})^2 + (u + \frac{a_2}{a_0 + a_3})^2\Bigr), \\
    -2\; X \cdot X_3 = (b_0 + b_3) \Bigl((t + \frac{b_1}{b_0 + b_3})^2 + (u + \frac{b_2}{b_0 + b_3})^2\Bigr), \\
    -2\; X \cdot X_4 = (c_0 + c_3) \Bigl((t + \frac{c_1}{c_0 + c_3})^2 + (u + \frac{c_2}{c_0 + c_3})^2 + \frac{c^2}{(c_0 + c_3)^2}\Bigr).
\end{gather}

Then, we can write the integral as
\begin{equation}
    J(n_a, n_b, n_c) = \int_{\mathbb{R}^2} \frac{d t d u}{(X \cdot X_1)^{2 - n_a - n_b - n_c} (X \cdot X_2)^{n_a} (X \cdot X_3)^{n_b} (X \cdot X_4)^{n_c}}.
\end{equation}

This is a box integral with two massive $X_1 \cdot X_1 \neq 0 \neq X_4 \cdot X_4$ and two massless $X_2 \cdot X_2 = X_3 \cdot X_3 = 0$ propagators, so we can write $J(n_a, n_b, n_c) = K(2 - n_a - n_b - n_c, n_a, n_b, n_c)$.  It can be further rewritten as an integral in the embedding space $X$
\begin{equation}
    K(n_1, n_2, n_3, n_4) = \int \frac{\delta(X^2) d^4 X}{\operatorname{Vol}(\mathbb{R}^{\times}) \prod_{i = 1}^4 (X \cdot X_i)^{n_i}},
\end{equation}
where $n_1 + \cdots + n_4 = 2$ and we quotient by the rescaling symmetry $X \to \lambda X$ which appears in this case.\footnote{This assumes the transformation $\delta(\lambda x) = \lambda^{-1} \delta(x)$, for all non-vanishing $\lambda$.  In particular, we do not have an absolute value of $\lambda$.  This notation is often used in this sense, but it would be better to use a residue with respect to $X^2 = 0$, which has the correct holomorphic rescaling in $X$.}

Recall that $a = 2 x_{12}$, $b = 2 x_{14}$, $c = 2 x_{13}$.  Using $X_1 \cdot X_i = -X_i^0$ we can see that in the notation of sec.~\ref{sec:squared-tadpoles}, $A_0 = 2 x_{12}^0 = -X_1 \cdot X_2$, $B_0 = -X_1 \cdot X_3$ and $C_1 = -X_1 \cdot X_4$.  We also have $C_0 = -X_4 \cdot X_4 - m^2$.

The integral of the double tadpole residue from sec.~\ref{sec:squared-tadpoles} becomes,
\begin{multline}
    \frac{1}{2} K(0, 1, 1, 0) 
    -\frac{1}{2} K(0, 2, 1, -1)
    -\frac{1}{2} K(0, 1, 2, -1) \\
    + \frac{1}{x_{13}^2 - m^2} \Bigl(
    -\frac{X_1 \cdot X_2}{4} K(1, 2, 1, -2)
    -\frac{X_1 \cdot X_3}{4} K(1, 1, 2, -2) \\
    +\frac{X_1 \cdot X_4}{2} K(1, 1, 1, -1)
    +\frac{1}{4} K(2, 1, 1, -2)
    \Bigr),
\end{multline}
where we neglect a global multiplicative factor of $C_0^{-2}$.
Taking the same residue for the integral $K(2, 1, -1, 1)$ we find
\begin{equation}
    -4 \frac{d t d u}{(1 + t^2 + u^2)^2} \Bigl(\frac{2}{A_1 B_1} - \frac{2 C_2}{A_1^2 B_1} - \frac{2 C_2}{A_1 B_1^2}\Bigr).
\end{equation}
This can be rewritten as
\begin{equation}
    -2 K(0, 1, 1, 0)
    +2 K(0, 1, 2, -1)
    +2 K(0, 2, 1, -1)
\end{equation}

The residues of the other integrals can be computed in a similar way.

%% file: nonplpent.tex
\section{Non-planar pentagonal two-loop finite integrals}
\label{sec:nonplpent}
In this appendix, we briefly describe the method used for the determination of the finite integrals for the non-planar pentagon topology in sec.~\ref{sec:two-loop-example}. The algorithm will be described in detail in future work. Here we summarize the steps involved:
\begin{itemize}
    \item \textbf{Identifying pinches}: Given a topology, we determine the set of relevant pinches by selecting the appropriate sub-topologies appearing in its contracted diagrams, according to the principle described in sec.~\ref{sec:intro}. Pinches which contribute to divergences are selected through purely diagrammatical rules.  
    \item \textbf{Cancellation of divergences}: Here we follow a similar procedure as in~\cite{Gambuti:2023eqh}; we introduce a basis of Lorentz-invariant products in terms of van Neerven-Vermaseren vectors $v_a$, defined such that $v_a\cdot p_i=\delta_{ai}$. This has the advantage of making the distinction between integrals that vanish in the limit $\epsilon\rightarrow0$ (evanescent integrals), and genuinely finite integrals manifest, as well as simplifying the form of expressions appearing when enforcing the cancellation of pinches. We define additional bases of higher order in loop momenta through products of these elements. Starting from the basis elements of linear order, we define a numerator as a generic combination of these. We evaluate the residue of the integrand expressions over finite fields for different values of the loop and external kinematics on the pinch surface. We repeat the procedure for each pinch surface. 
    \item \textbf{Characterizing a basis}: We solve the resulting system and we reconstruct the symbolic expression for numerators. Having identified the solution at this order in the loop momenta, we can immediately determine that all terms given by their product with additional monomial will be finite as well, as long as it does not saturate the bound on the power of loop momenta in numerators imposed by UV-finiteness by the conditions of Weinberg's theorem \cite{Weinberg:1959nj}. We then proceed to monomials of higher orders, until we reach the bound determined by UV-finiteness, at each step removing those terms corresponding to products containing lower-order solutions from the space of numerators, in order to identify independent finite-numerator generators that can arise.
\end{itemize}
In terms of the routing in fig. \ref{fig:double-pentagon}, we can describe the pinches whose cancellation we need to impose by the subset of propagators contained in the contracted diagram associated to each of them. These are
\begin{align}
   & \{q_1,q_2\},\{q_4,q_5\},\{q_3,q_4\},\{q_1,q_6\},\{q_7,q_8\} \qquad&\text{ at one loop}\\~\nonumber\\
   &\nonumber\{q_1,q_2,q_5,q_7\},\{q_1,q_2,q_4,q_5\},\{q_1,q_2,q_3,q_4\},\\
   &\nonumber\{q_1,q_6,q_3,q_8\},\{q_1,q_6,q_4,q_5\},\{q_1,q_6,q_3,q_4\},\\
   &\nonumber\{q_4,q_5,q_2,q_7\},\{q_3,q_4,q_6,q_8\},\{q_7,q_8,q_2,q_5\},\\
   &\nonumber\{q_7,q_8,q_3,q_6\},\{q_7,q_8,q_1,q_2\},\{q_7,q_8,q_1,q_6\},\\
   &\{q_7,q_8,q_3,q_4\},\{q_7,q_8,q_4,q_5\}\qquad \qquad &\text{at two loops}
\end{align}
Results for all non-evanescent integrals are available in the ancillary files, both in terms of generators and of the full basis elements up to order seven in the loop momenta. They are expressed as a list of numerator polynomials in terms of products with the vectors $v_a$ defined above.